\documentclass[12pt]{article}
\usepackage[latin1]{inputenc}
\usepackage{color}
\usepackage{natbib}
\usepackage{graphicx}

\definecolor{vert}{rgb}{0,1,0}
\definecolor{rouge}{rgb}{1,0,0}
\definecolor{bleu}{rgb}{0,0,1}
\newenvironment{narrow}[3]{%
\begin{list}{}{%
\setlength{\topsep}{0pt}%
\setlength{\leftmargin}{#1}%
\setlength{\rightmargin}{#2}%
\setlength{\topmargin}{#3}%
\setlength{\listparindent}{\parindent}%
\setlength{\itemindent}{\parindent}%
\setlength{\parsep}{\parskip}}%
\item[]}{\end{list}}

\begin{document}

\def \indi{\hbox{1}\!\hbox{I}}
\begin{center}
{\Large \sc Stable variable selection for right censored data: comparison of methods}

\bigskip 
Marie Walschaerts\footnotemark 
, Eve Leconte\footnotemark   \& Philippe Besse\footnotemark

\medskip
{\it
$^1$Equipe d'accueil EA3694, Recherche en Fertilité Humaine,\\
 Hôpital Paule de Viguier, 330 avenue de Grande-Bretagne, 31059 Toulouse\\
$^2$TSE (GREMAQ), Universit\'e Toulouse 1 Capitole,\\
  21 all\'ee de Brienne,  31000 Toulouse\\
$^3$IMT, UMR CNRS 5219, Université de Toulouse, INSA \\
118 route de Narbonne, 31062 Toulouse Cedex 9  \\
}
\medskip
E-mail : walschaerts.m@chu-toulouse.fr, leconte@cict.fr, philippe.besse@math.univ-toulouse.fr \\
\end{center}

\begin{center}
{\large Abstract}
\end{center}
The instability in the selection of models is a major concern with
data sets containing a large number of covariates. This paper deals
with  variable selection methodology in the case of high-dimensional
problems where the response variable can be right censored. We focuse
on new stable variable selection methods based on bootstrap for two
different methodologies commonly used in survival analysis: the Cox
proportional hazard model and survival trees. As far as the Cox model is concerned, we investigate the bootstrapping applied to two variable selection techniques: 
the stepwise algorithm based on the AIC criterion and the  
$\mathcal{L}_1$-penalization of Lasso. 
Regarding survival trees, we review two methodologies: the bootstrap node-level stabilization and random survival forests.
We apply these different approaches to two
real data sets, a classical breast cancer data set and an original
infertility data set.  We compare the methods on two criteria: the prediction error rate  based on the Harrell concordance index and the relevance of the interpretation of the corresponding selected models, focusing on the  original
infertility data set.
The aim is to find a compromise
between a good prediction performance and ease to interpretation for
clinicians.
Results suggest that in the case of a small number of
individuals, a bootstrapping adapted to $\mathcal{L}_1$-penalization
in the Cox model or a bootstrap node-level stabilization in survival
trees give a good alternative to the random survival forest
methodology, known to give the smallest  prediction error rate but difficult to interprete 
by non-statisticians. In a clinical perspective, the complementarity between
the methods based on the Cox model and those based on survival trees
would permit to built reliable models easy to interprete by the clinician.

\vspace{0.5cm}\noindent {\bf Key-words:} censored data, variable selection, survival trees, survival random forests, Lasso, Cox model, bootstrap.

\newpage
\section{Introduction}
Problems of variable selection arouse a growing interest in the
processing of data sets containing more and more variables.  In the
last twenty years, many methods of variable selection have been
proposed to handle these high-dimensional problems, especially when
the number of covariates $p$ exceeds the number of observations $n$.
To avoid a wrong estimation due to collinearity problems and to
improve interpretation, the scientific community has developed tools
to select the most relevant variables. A large literature concentrates
on the case of the linear regression. 
A classical well-known method is the
stepwise algorithm based on the Akaike Information
Criterion (AIC). Recently, another field of research has focused on
optimization problems, such as $\mathcal{L}_1$-penalty approaches.
On
the other way, tree-based algorithms provide an interesting
alternative to handle non-parametrically a large number of covariates.

  We consider the special case where the response variable is
 right censored.  In this context, the Cox proportional hazards model [1] has
 become the gold-standard tool for the statistical analysis,
 especially in the medical field. However, in the case of a large
 amount of covariates, it may be very unstable, even when
 stepwise selection  or $\mathcal{L}_1$-penalization of Lasso
 are added to the classical procedure.  
The instability of these selection approaches have also been encountered  in the context of the linear regression [2, 3]. To
 remedy this problem, some authors  have proposed to use bootstrapping
 to investigate the reliability of the  choice of the variables in the final model.
 Bach [4] introduced stability in the selection by using bootstrapping in a Lasso  algorithm, a method called Bolasso, and Meinshausen and Bühlmann [3] proposed to improve the $\mathcal{L}_1$-penalization by randomizing the selection process of the covariates. 
These  bootstrapped Lasso methods have only been considered in the framework of  linear regression: we propose to extend them to the stabilization of the selection of covariates in a Cox model.

Alternatively to the Cox model, survival trees procedures permit to
take into account non linear relationships between the censored
variable and the covariates and yield easily interpretable
classification rules to the clinicians, but they tend to overfit the
data and suffer also from instability [5 -- 8].
A stabilization technique
known to improve the prediction performance of a single tree consists
in aggregating a family of tree models using boostrapped samples and a
random selection of the covariates at each node of the trees. This
procedure, called ``random forests'' [9] has been adapted to the
survival framework by Ishwaran \textit{et al.} [10]. This Random
Survival Forest (RSF) methodology is considered as the best
modelization in terms of prediction performance but it it is not easy to
interprete as it does not provide a single tree. On the other way, Dannegger [6]
introduced tools to control the stability of the selection of the
covariates at each node of a tree and propose resampling techniques at
node-level to stabilize the choice of the split. This methodology has
the advantage to keep the simplicity of interpretation of a single
final tree.

We review and compare these different stable variable selection
methods derived on the Cox model or on survival trees on real data sets.
 We will base the comparison of the proposed methods on statistical
 criteria as the prediction accuracy but also on the capacity of the
 method to supply a final model whose clinical interpretation
 is easy and pertinent.  As a matter of fact, the objective of the
 statistician is to provide to clinicians a model which is robust
 enough to permit a proper understanding of the relationships between
 the variable of interest and the covariates in  a predictive and explanatory purpose.

We consider two real data sets.
 The first  data set concerns the survival of breast cancer patients in relation with their gene-expression signature [11] and  is a very usual
case encountered in survival analysis. This classical example will validate the
different approaches discussed earlier. The second data set is much more original  as it concerns couples consulting for infertility and very few studies were led in this domain.
The censored variable of interest is the duration between the first visit of the couple and the birth of a alive child, which is difficult
to model. The study takes into account several covariates of the couple or  man and woman, but also the associated medical treatment according to the
diagnosed  causes of infertility. For example, a man with severe oligospermia
will receive ICSI (IntraCytoplasmic Sperm Injection), while a woman
with a dysovulation or with no tubal factor will enter an IVF (In
Vitro Fertilization) program. In many couples,  causes of
infertility are not identified and physicians can not clearly answer
the question of on-going pregnancy in couples.

In section 2,  a presentation of the  variable selection methods based on the Cox model is given, followed, in section 3, by a review of the selection methods based on trees algorithms. Section 4 compare the different approaches on the two data sets,  giving the prediction erro rate and the selected variables in the final model for each approach. Concluding remarks  and perspectives are presented in the last section.

\section{Stable variable selection methods based on the Cox model}

\subsection*{Notations} 
We denote by $T$ the variable of interest, which is a time of
failure. We suppose that $T$ may be right censored at a
non-informative censoring time $C$ such that $C$ is independent of $T$
conditionally on $Z$, a $p$-vector of covariates. The observations
are therefore the possibly censored time $X=\min(T,C)$ and the
censoring indicator $\Delta=1\{T\leq C\}$, where $1\{\cdot\}$ is the
indicator function. 
The
observed sample is thus $(X_i,\Delta_i,Z_i)_{i=1,\ldots,n}$.

In survival analysis, the Cox model [1] has become the most popular method to modelize the relationship between a survival time and one or more predictors (\textit{i.e.} covariates).
This model  has the advantage to be semi-parametric in the sense that it does not require assumptions on the survival time distribution. Morevoer, it is of easy interpretation for clinicians in providing estimates of the effect of the covariates on survival time  after adjustment on the other covariates. 

In the Cox regression model in its simplest form, the hazard function for the failure time of an individual takes the form
$$ \lambda(t|Z)=\lambda_{0}(t)\exp(\beta'Z),$$
where  $\beta$ is a $p$-vector of unknown regression parameters and $\lambda_{0}(t)$ is an unknown baseline hazard function.  Denote $t_{(1)},\dots,t_{(k)}$  the $k$ ordered uncensored survival times. The parameter of interest $\beta$ is estimated by maximising the logarithm  of the partial likelihood function
\begin{eqnarray}\label{partlik}
L(\beta) = \sum_{i=1}^k \left( \beta' Z_i - \log \left(\sum_{j
\in {\cal R}_i} e^{\beta' Z_j}\right) \right)
\end{eqnarray}
where $R_i$ is the set of indices of the individuals at risk at time $t_{(i)}$.
Notice that the Cox model is a log-linear model based on the proportional hazard hypothesis, which stipulates that the ratio of the hazard functions of two individuals is constant over time.\\

In the following, we review three stable methods to select the relevant covariates using the Cox model: the bootstrap stepwise selection, the bootstrap Lasso selection and the bootstrap randomized Lasso selection.

\subsection{Bootstrap stepwise selection (BSS)}
In case of a large number of covariates, the selection of the
predictors in the Cox model is usually made by a stepwise algorithm
minimizing the Akaike Information Criteron (AIC).
However, the stability of this method is questionable.   Using a data splitting
approach, Harrell \textit{et al.} [2]
showed variation in the selected predictors. This result was confirmed by Chen and George [12] who applied the
stepwise procedure on 100 bootstrapped samples from a study of acute
lymphocytic leukaemia. The original model which was built from the
entire set by a Cox regression model coincide in only 2\% of the cases with
the models selected from the bootstrapped samples.
Sauerbrei and Schumacher [13] developped a bootstrap selection procedure
which combined the bootstrap method with stepwise selection in Cox
regression. They examined the inclusion frequencies of the variables
selected by the stepwise algorithm into the models derived from the bootstrapped samples and
keep in the final model the variables for which the inclusion
frequency exceeds a given cut-off value $\kappa$ in $(0,1)$. The
choice of $\kappa$ is arbitrary. The authors applied the methodology
to a data set on brain tumours including 447 patients and 12
covariates, using two values of $\kappa$ ($0.3$ and $0.6$). They
showed that the choice of $\kappa=0.6$ gives the same final model
whatever the number of bootstrap samples (from $100$ to $1000$).

\subsection{Bootstrap Lasso and bootstrap randomized Lasso selection}

\subsubsection*{Lasso selection}
An alternative to the stepwise selection procedure is the Lasso
selection. Adapted by Tibshirani [14] to the Cox model, the
method estimates the $\beta$ parameter via maximising the log partial
likelihood function (\ref{partlik}) with the constraint
$\sum_{j=1}^p\left|\beta_j\right| \leq \lambda$ where $\lambda$ is a
regularization parameter. The Lasso constraint selects variables by
shrinking estimated coefficients towards 0. This leads to coefficients
exactly equal to zero and allows a parcimonious and interpretable
model.  The choice of a proper $\lambda$ is sensitive and leads to
variation in model selection as demonstrated by
Meinshausen and Bühlmann [3]. In a linear regression framework, they studied
the stability of the selection in a gene expression data set. They introduced
noise in variables in permuting all but 6 of the 4088 gene
expressions. They showed that, when $\lambda$ increases, the final
model retained as well the 6 unpermuted variables as the irrelevant
noised variables.  Although cross-validation seems a natural solution
to choose $\lambda$, it is not a good alternative for high-dimensional
problems. As a matter of fact, the authors showed in their example
that 14 permuted variables are still retained in the final model obtained in
choosing $\lambda$ by cross-validation. Moreover,
Meinshausen and Bühlmann [15] prooved that if the number of covariates tends
to infinity then the probability of selecting a wrong variable
converges to $1$.

\subsubsection*{Bootstrap Lasso selection (BLS)}
In order to obtain a consistent model selection, 
 Bach [4] proposed for the linear model a bootstrapped version of the Lasso,
 referred as the Bolasso. He defined the Bolasso model estimate as the
 final model composed by only the variables which are selected in all
 bootstrapped samples; in other words, using the terminology of the BSS, it corresponds to a cut-off
 value $\kappa$ equal to 1. However, 
Meinshausen and Bühlmann [3] showed with the permuted gene data set that even with random subsampling
 (which is a procedure close to the bootstrap [16]), 
 the Lasso algorithm could select irrelevant variables when $\lambda$
 is too large. 

\subsubsection*{Bootstrap randomized Lasso selection (BRLS)}
To deal with the choice of $\lambda$, 
Meinshausen and Bühlmann [3] proposed a generalisation of the bootstrap Lasso
procedure called bootstrap randomized Lasso where the covariates are penalized by different values randomly chosen in the range $[\lambda,\lambda/\alpha]$ with $\alpha$
in $(0,1)$. This turns out to estimate the $\beta$ parameter with the constraint $\displaystyle \sum_{j=1}^p \left|\frac{\beta_j}{W_j}\right| \leq \lambda$. In practice, the set of covariates
 $\{Z_j:\ j=1,\ldots,p\}$ are weighted by the set
 $\{W_j:\ j=1,\ldots,p\}$ randomly generated where $P(W_j = \alpha) = p_w$ and $P(W_j = 1) = 1-p_w$ with $p_w$  in $(0,1)$. The authors give no indication on the choice of $p_w$  and in the following we take $p_w=0.5$.
This procedure is very simple to implement and the authors showed that choosing $\alpha$
in the range of $(0.2,0.8)$ gives a consistent
variable selection whatever the choice of the penalty $\lambda$. The authors explained that a low value of $\alpha$ will decrease the selection probabilities of irrelevant variables even if the penalty $\lambda$ is large.
As the BLS and BRLS methods are based on  a penalization of the parameter $\beta$, it is necessary to first normalize the covariates  such as $|| Z_j||_2 = (\sum_{i=1}^n Z_{ij}^2)^{1/2} = 1$ for all $j$ in $\{1,\ldots,p\}$.

\section{Stable variable selection methods based on survival trees}

Although they are not so popular than the Cox model, tree-based methods in survival analysis (the so-called survival trees) have known a great development in the last decades.  They provide a good alternative to the Cox regression model in identifying covariates which play a role on the survival outcome and in predicting the individual risk of failure. In addition to be easy to interpret in a large frame of applications, survival trees methods can incorporate non linear effects, and also take into account interactions between covariates. 

First developed for basic classification trees, the Classification and Regression Trees (CART) algorithm of Breiman [17] is based on binary recursive partitioning. This is an iterative process which splits the data into two subgroups (daughter nodes) according to the value of one of the predictors. The splitting rule maximises the difference between nodes. Let denote $\{Z_j:j=1,\dots,p\}$ the set of covariates. Formally, a split is induced by a question of the form ``Is $Z_j \leq b$ ?'' for a continuous covariate where $b$ is a cut-off value in the set of realizations of $Z_j$. For nominal covariates with possible values in the set $B =\{b_1,\ldots,b_r\}$, the question is of the form ``Is $Z_j$ in $S$?'' where $S \subset B$. To determine the best split $s$ among all covariates in the current node $h$, a measure of improvement $G(s,h)$ is evaluated, producing the most homogeneous daughter nodes. The best split $s_{best}$ is the optimal split among all possible splits in the set $S_h$ such as 
$G(s_{best},h) = max_{s \in S_h}G(s,h)$.
 The process goes on until each node reaches a user-specified minimum node size and becomes a terminal node, or is homogeneous.
 To control the size of the tree, a stopping rule is used to prune the ``large'' tree containing ``pure'' nodes.
The method employed by CART is called cost-complexity pruning. The complexity of a tree $\Theta$ is : 
$$R_{cp}(\Theta) = R(\Theta) + cp \left|\tilde{\Theta}\right|,$$
where $R(\Theta)$ is the raw error of  measure (sum of the error of measure over the terminal nodes), $\left|\tilde{\Theta}\right|$ is the number of terminal nodes, and $cp$ is an arbitrary penalty weight between 0 and $\infty$ called the complexity parameter. 
A sequence of nested trees is built. The final tree is the smallest tree for which $R_{cp}(\Theta)$ is minimized.
To deal with the choice of $cp$, cross-validation techniques can be used to determine the optimal size of the  tree.

Various splitting and pruning approaches have been proposed to adapt regression trees to survival data. Gordon and Olshen [18] used the Wasserstein metrics to measure the distance between two Kaplan-Meier estimates of the survival distribution. The split criterion chooses the predictor (and if necessary the cut-off) which maximises this distance for the left and right daugther nodes.
To prune the survival tree, they generalized the cost-complexity measure for censored data. 
A more usual way to measure the difference between survival curves is the logrank test statistic. This was done  by Ciampi \textit{et al.} [19], Segal [20] and LeBlanc and Crowley [21] who suggested to use the logrank test statistic as a between-node heterogeneity measure. As an alternative to the standard cost-complexity pruning approach using proportional hazards martingale residuals as error measures, LeBlanc and Crowley [21] proposed a ``goodness-of-split'' complexity based on the sum of the standardized splitting logrank test statistics of the internal nodes.
Others authors suggested to use split criteria based on the likelihood function. Davis and Anderson [22] assume that the survival function within each node is an exponential function with a constant hazard, while LeBlanc and Crowley [23] only assume proportionality for the hazard functions of two daughter nodes. These latter used for estimation the full or partial likelihood function in the Cox proportional hazards model.

Instability in the selection of covariates by regression trees has
been observed and demonstrated by many authors [5 -- 8] 
This instability may be due to an overfitting of data. The variance observed may
also come from arbitrary cutpoints defined by the dichotomization of
continuous covariates. We review in the following two remedies to
instability of survival trees: the bootstrap node-level  stabilization
proposed by Dannegger [6] and the random survival forests of
Ishwaran \textit{et al.} [10].

\subsection{Bootstrap node-level stabilization (BNLS)}

Dannegger [6] proposed a bootstrap node-level stabilization procedure for survival trees. The algorithm consists, at node $h$, in drawing bootstrapped samples from the original set, and for each of them, in finding the best split. The split which appears the most of the time at the node $h$ is selected. For a continuous variable, the cut-off value $b$ chosen in the set of realizations of the split variable is the median of all the $b$-values proposed at each bootstrap. As Dannegger [6] did not propose a choice of the cut-off for a categorical variable, we decide to
affect to the daughter node the level of the  categorical variable which was  mainly chosen  by the boostrapped samples.
To find the optimal complexity parameter $cp$ to prune the tree, we use  a ten-fold cross-validation procedure as suggested by Dannegger [6].

Using simulated data, Dannegger [6] compared his methodology (for 100 bootstraps at each node) to the  CART algorithm and to the bagging method and found that the  BNLS reduces the predictor error rate from 26\% (the value obtained with CART) to 6\%. The bagging method has the best performance  with a prediction error  of 3\%. However the model obtained with this latter method is not easy to use and interpret by non-statisticians, unlike BNLS.

\subsection{Random Survival Forests (RSF)}

In order to stabilize the trees obtained by the CART procedure, Breiman [5] proposed the bagging method (so called for bootstrap aggregating) based on a family of random trees: multiple versions of predictor models obtained from bootstrapped samples are aggregated in order to construct a robust model estimator. Bagging was then adapted to the survival framework by Efron [24] and Akritas [25]. 
Another way to improve stability of trees is the boosting, developped by Freund and Schapire [26].
As for the bagging, the boosting consists in aggregating a family of models. Each tree is built iteratively from a weighted sample (an individual who is misclassified gains weight and an individual who is classified correctly looses weight)  and then evaluated according to its ability to classify data. Considered as better than the bagging, the boosting is yet limited when data are too noised:
the algorithm gives a heavy weight to noised data which leads to a bad overfit [27].

Breiman [5] proposed a random selection approach which
combines the bagging method with a random selection of the covariates
at each node of the tree. This methodology, called random forests, is
more stable than the two previous ones [5, 26].
The method was adapted to the survival framework
in an approach called ``random survival forests'' by
Ishwaran [10]. Bootstrapped samples were drawn from the original
data set. Notice that each bootstrapped sample excludes on average
37\% of the data, a set called out-of-bag (OOB) data.  For each
bootstrapped sample $b$ in $(1,\ldots,ntree)$, a survival tree is
built: at each node, a subset of covariates is randomly selected among
all the covariates. The splitting process continues under the
constraint that each terminal node contains at least a fixed number of
unique event times. The RSF algorithm computes an ensemble estimate
for the cumulative hazard function (CHF), which is used as a
predictor.  For each terminal node $h$, let
$t_{(1,h)}<\ldots<t_{(N_h,h)}$ be the distinct ordered uncensored event
times, and define $d_{l,h}$ and $Y_{l,h}$ as the number of events and
individuals at risk at time $t_{(l,h)}$, for $l=1,...,N_h$ respectively. 
The CHF
estimator for the node $h$ is the Nelson-Aalen
estimator $$\hat{H}_h(t) = \sum_{t_{(l,h)}\leq t}
\frac{d_{l,h}}{Y_{l,h}}\cdot$$ 
Each tree provides such a CHF estimate at 
each terminal node. Let $\hat{H}_b(t|z)$ denote the cumulative hazard
estimate for tree $b$ conditionnaly on the covariate $z$.  To determine the CHF
estimate for individual $i$ with covariate vector $z_i$ obtained from the
tree $b$, drop $z_i$ down the tree. It will fall in a unique terminal node $h$. So we have  $\hat{H}_b(t|z_i) = \hat{H}_h(t)$ if $z_i \in h$. 
Let $I_{i,b} = 1$ if
$i$ is an OOB point for $b$ and 0 otherwise.  The OOB ensemble
CHF  estimator for $i$ is: 
\begin{equation}
\hat{H}_e^*(t|z_i) =
\frac{\sum_{b=1}^{ntree}I_{i,b}\hat{H}_b(t|z_i)}{\sum_{b=1}^{ntree}I_{i,b}}\cdot
\label{OOBCHF}
\end{equation}
Note that  the above  estimator is
obtained by avering  only over  bootstrap samples in which $i$ is
excluded.

\section{Comparison of the methods}

In order to compare the different variable selection approaches, we apply them to two real data sets, the first concerning breast cancer and the second male fertility. The five following procedures have been compared: the bootstrap Cox stepwise procedure (BSS), the bootstrap Cox Lasso procedure (BLS), the bootstrap Cox randomized Lasso procedure (BRLS) with three different values of $\alpha$ (0.2, 0.4 and 0.6), the bootstrap node-level stabilization procedure (BNLS) and the random survival forest method (RSF).

\subsection{Software}

We use the methods implemented in {\tt R} software (CRAN). We adapted
the bootstrap to the stepwise Cox algorithm and to the {\tt R} package
{\tt penalized}. We also modified the {\tt R} package {\tt rpart} in
order to introduce bootstrap at node-level in the building of the tree
(BNLS procedure).  Instead of the default {\tt exp} method which
maximises an exponential likelihood, we used as a splitting criterion
in {\tt rpart} the logrank test statistic, following the
recommandations of Radespiel-Troger \textit{et al.} [28]. They showed
that the prediction error of the tree is related to the instability of
the covariate selection at each node. Compared to other splitting
criteria on a real survival data set on gallbladder stones, the
logrank test statistic yields the lowest prediction error (evaluated
by the Brier score). The {\tt R} package {\tt randomSurvivalForests}
provides four different splitting rules: logrank, conservation of
events, logrank score, and approximate logrank (for more details, see
Ishwaran and Kogalur [29]). We used the default splitting criterion
\textit{i.e.} the logrank test statistic.  Concerning the number of
bootstrapped samples, we take $N=100$ for the Cox model based
procedures, the default value $N=1000$ for RSF, and $N=1000$ for the
BNLS procedure.

\subsection{Comparison criteria}

We compare the five procedures on the basis on their prediction performance but also on the usefulness of the
model for non-statisticians.
The prediction performance is measured by the prediction error rate. A model will be
considered as useful if the statistical method involved is the most
adapted to the problematic of the clinician and if the results are easy to interprete.

\subsubsection{Comparison of the prediction error rates}

We compare the stability of the five procedures using a statistical criterion: the prediction error rate, which allows to quantify the 
prediction performance of the final model selected.
To this aim, the original data set is divided into two subsets, the training set and the
test set. The procedures, computed on the training set, give a
final model which is then applied on the test set to calculate the
prediction error rate. We obtain the  prediction error rates in three steps:
\begin{enumerate}
	\item We graphically chose sensible values of the cut-off $\kappa$ for the methods based on the Cox model and also values of $\lambda$ for BLS and BLRS. For that, considering the whole set, we plot the values of $\kappa$ for each selected covariable for BSS and we  plot the values of $\kappa$ with respect to values of $\lambda$ for BLS and BLRS. 
The aim is to identify the most relevant variables whose frequencies of inclusion $\kappa$ are larger than those of the other covariates whatever the value of $\lambda$. Then we chose values of $\lambda$ and $\kappa$ which maximize the gap between the two subgroups of covariates.
For BNLS, the choice of the optimal complexity parameter $cp$ is achieved by cross-validation techniques.
	\item We then apply the procedures on the training set with the values of $\lambda$, $\kappa$ and $cp$ chosen at step 1. As far as the methods based on the Cox model are concerned, for each method, we adjust on the training set a simple Cox model including the subset of covariates for which the inclusion frequencies are greater than $\kappa$. This permits  to obtain the estimations of the $\beta$ parameter in order to calculate the predicted outcomes.
	\item We finally calculate the prediction error rates on the test set for each method.
\end{enumerate}

The most commonly used prediction error rate for survival models is based on the Harrell's
concordance index $C$ [30]. To compute the $C$-index,
 the observed and predicted outcomes are compared, as follows:
\begin{itemize}
	\item Form all possible pairs of observations.
	\item Eliminate those pairs where the  shorter survival time is censored, and also pairs where the survival times and the censoring indicators are equal.
	\item For each permissible pair, count $1$ if the shorter survival time has the worse predicted outcome and count $0.5$ if the predicted outcomes are tied. Sum over all permissible pairs.
	\item The C-index is the ratio of the latter sum  over the number of permissible pairs.
\end{itemize}
The prediction error rate is defined as $1-C$ and is in $\{0,1\}$. Note that a value of $0.5$  is not better than  random guessing.

For the Cox model based procedures, the prediction for an individual $i$  with covariate vector $Z_i$
is based on the linear predictor $\hat{\beta}'
Z_i$. Individual $i$ has worse predicted  outcome than $j$ if $\hat{\beta}'
Z_i > \hat{\beta}'Z_j$.
The predictions of the survival trees based procedures are derived from the CHF estimator. For the RSF procedure, they are derived from the OOB ensemble CHF estimator of formula (\ref{OOBCHF}).   Let $t_1, \ldots, t_m$ be the unique  times in the test set. Individual $i$ is said to have a worse predicted outcome than $j$ if $$ \sum_{k=1}^{m} \hat{H}_e^*(t_k|Z_i) > \sum_{k=1}^{m} \hat{H}_e^*(t_k|Z_j).$$

A total of $30$ training sets and test sets were
  drawn to obtain a sample of  error rates. Boxplots of the error rates are presented in order to compare the variability of the prediction performances.

\subsubsection{Comparison of the selected covariates}

We compare the set of covariates selected by the five different approaches applied on the entire set. For BSS, BLS and BRL, we take the values of the cut-off $\kappa$ and of the penalties $\lambda$ determined above. 
The BNLS algorithm provides the final tree which has been pruned with a fixed complexity parameter $cp$.
For the RSF procedure, the impact of the variables is measured by their ``variable importance'' (VIMP), which is the difference between the prediction error obtained  with the original RSF procedure and the prediction error obtained using randomized assignments whenever a split for the considered variable is encountered [10]. 

\subsection{Results for the breast cancer data set}

\subsubsection{Description}
The breast cancer data set  contains the metastasis-free survival times  from the study of Vijver \textit{et al.} [31]
 who classified a serie of 295  patients with primary
breast carcinomas as having a gene-expression signature associated with either a poor or a good prognosis. We restricted the study to the 144 patients who had lymph node positive disease. Vijver \textit{et al.} [31] evaluated the predictive value of the gene-expression profile of patients for the 70 genes previously determined by Veer \textit{et al.} [32] based on a supervised learning method. The data set can be found in the {\tt R} package {\tt penalized}. Five clinical risk factors and 70 gene expression measurements found to be prognostic for metastasis-free survival have been recorded. The censoring rate is 66\%.
The  variables in the data set are:
\begin{itemize}
	\item {\tt time}: metastasis-free follow-up time,
	\item {\tt event}: censoring indicator (1 = metastasis or death; 0 = censored),
	\item {\tt diam}: diameter of the tumor (two levels),
	\item {\tt N}: number of affected lymph nodes (two levels),
	\item {\tt ER}: estrogen receptor status (two levels),
	\item {\tt grade}: grade of the tumor (three ordered levels),
	\item {\tt age}: age of the patient  at diagnosis,
	\item {\tt TSPYL5}...{\tt C20orf46}: gene expression measurements of 70 prognostic genes.
\end{itemize}

\subsubsection{Prediction error rates}

The Cox stepwise algorithm did not converge for the BSS procedure, probably due to too many covariates and not enough events in the boostrapped samples.
As suggested by Meinshausen and Bühlmann [3], we search graphically the penalty $\lambda$  which leads to the best split between covariates in the BLS and BRLS procedures. However, the graphical determination of $\lambda$ is not so easy as in  Meinshausen and Bühlmann [3] which presents results on an ad-hoc example.
Figure $1$ suggests that selecting between four and five covariates seems to be the most relevant whatever the value of $\lambda$ and $\kappa$. 
So we decide to choose a value of $\lambda = 0.4$ and a value of $\kappa = 0.2$ for BLS and BRLS whatever the value of $\alpha$.
The complexity parameter $cp$ of the BNLS procedure is obtained by a ten-fold cross-validation procedure: we chose graphically  $cp=0.002$.

\begin{figure}
	\centering
		\begin{tabular}{cccc}
			\resizebox{8cm}{!}{\includegraphics{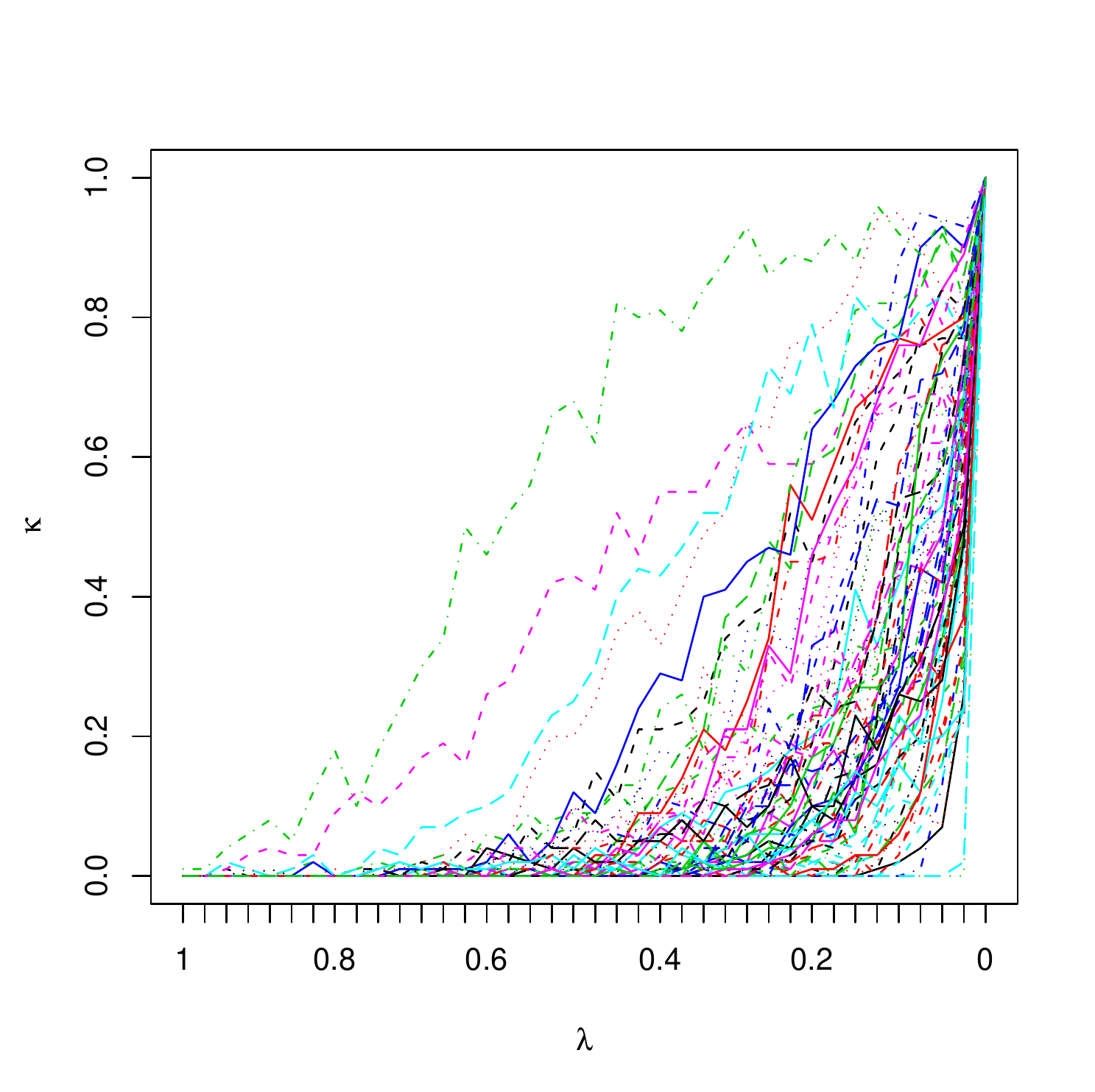}} & \resizebox{8cm}{!}{\includegraphics{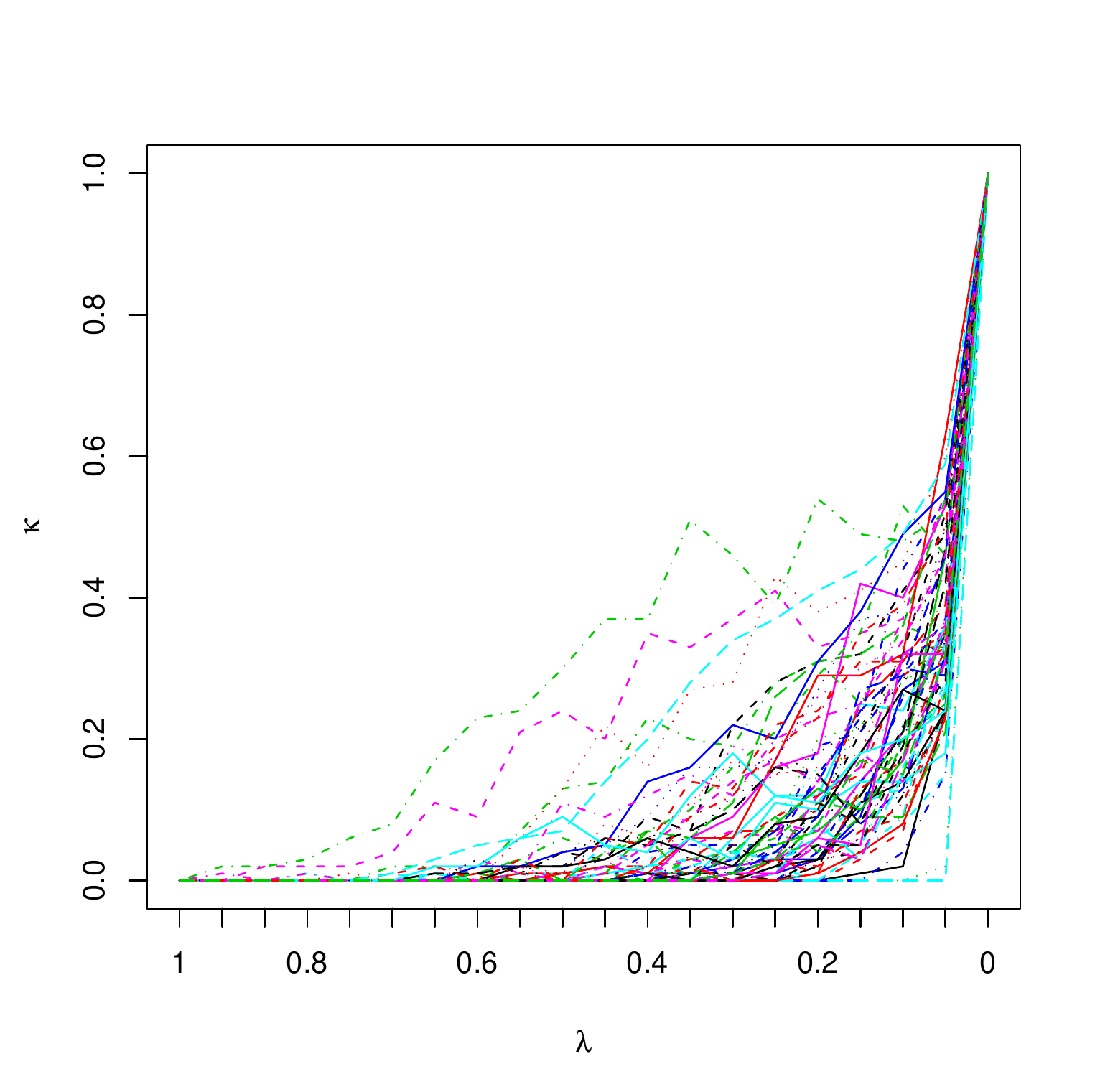}}\\
			\textit{1. BLS} & \textit{2. BRLS, $\alpha = 0.2$} \\
			\resizebox{8cm}{!}{\includegraphics{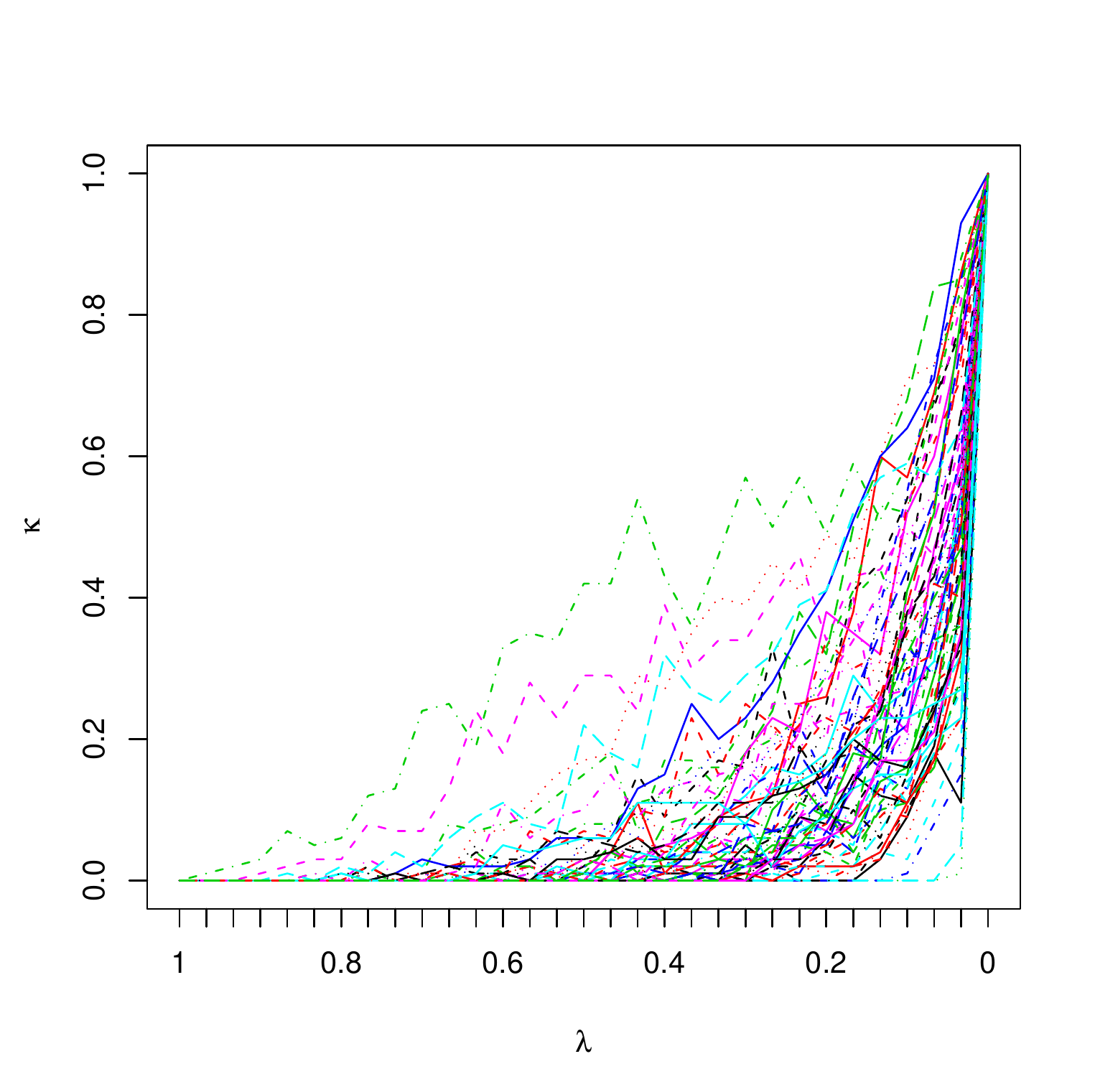}} & \resizebox{8cm}{!}{\includegraphics{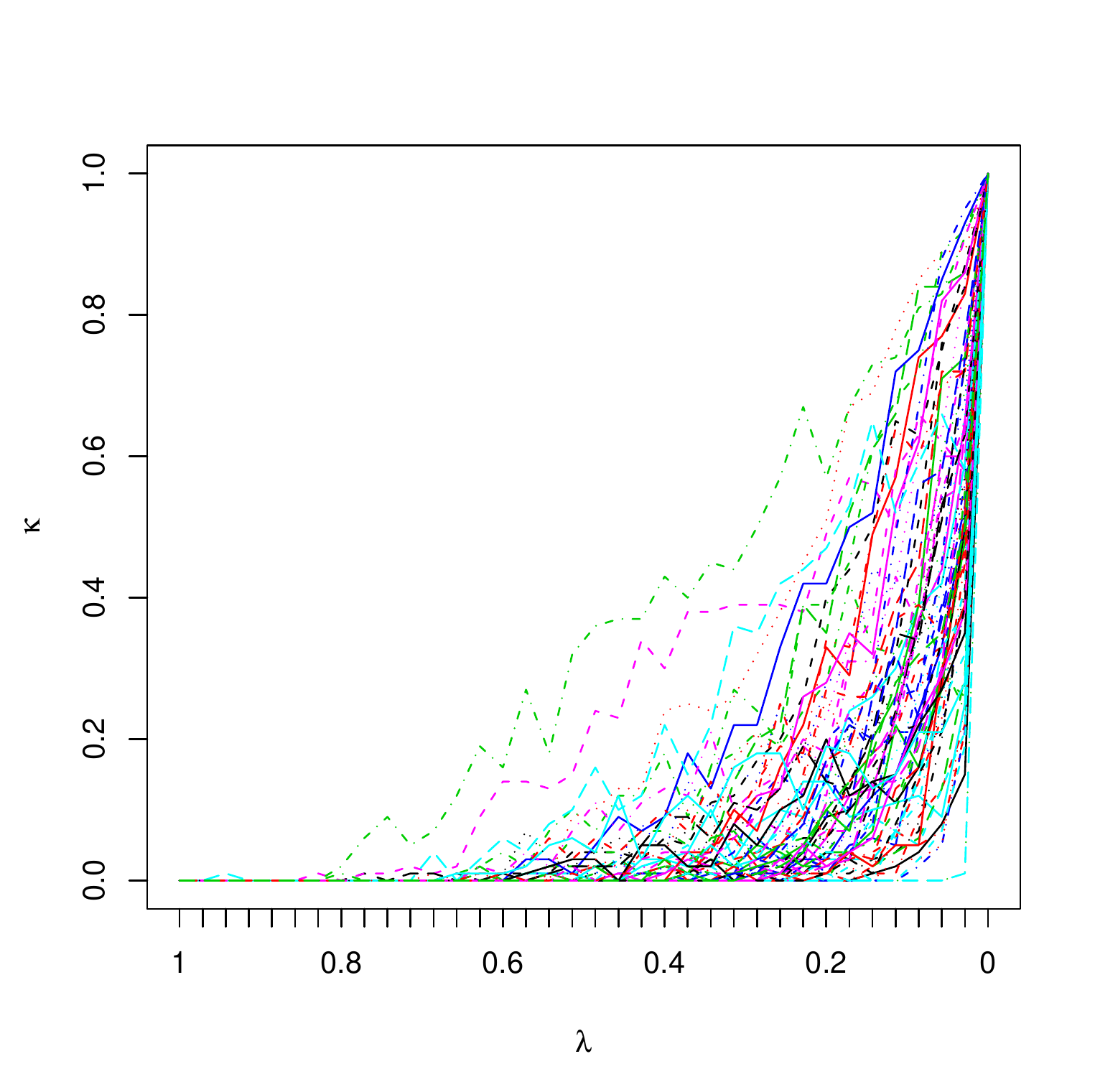}} \\
			\textit{3. BRLS, $\alpha = 0.4$} & \textit{4. BRLS, $\alpha = 0.6$} \\
		\end{tabular}
	\caption{Inclusion frequencies  $\kappa$ with respect to $\lambda$ for Bootstrap Lasso Selection and Bootstrap Randomized Lasso Selection.}
	\label{lasso_breast}
\end{figure}

Figure $2$ presents the boxplots for the different approaches based on 30 iterations and the numerical summaries are in table \ref{moy_breast}. As expected,  RSF gives the lowest mean,  median and standard deviation of the prediction error rates. We verify that the single survival tree has the worse prediction performance, but the BNLS procedure is hardly better.
 Whatever the value of $\alpha$,  the BRLS procedure shows a  large dispersion of error rates compared to the BLS procedure which has a standard deviation of $0.06$, similar to this of RSF. However, a good compromise between dispersion and mean of error rates is observed for BRLS with $\alpha = 0.2$ or $\alpha = 0.4$.
 
\begin{figure}
\centering
\vspace{-2cm}\resizebox{12cm}{!}{\includegraphics{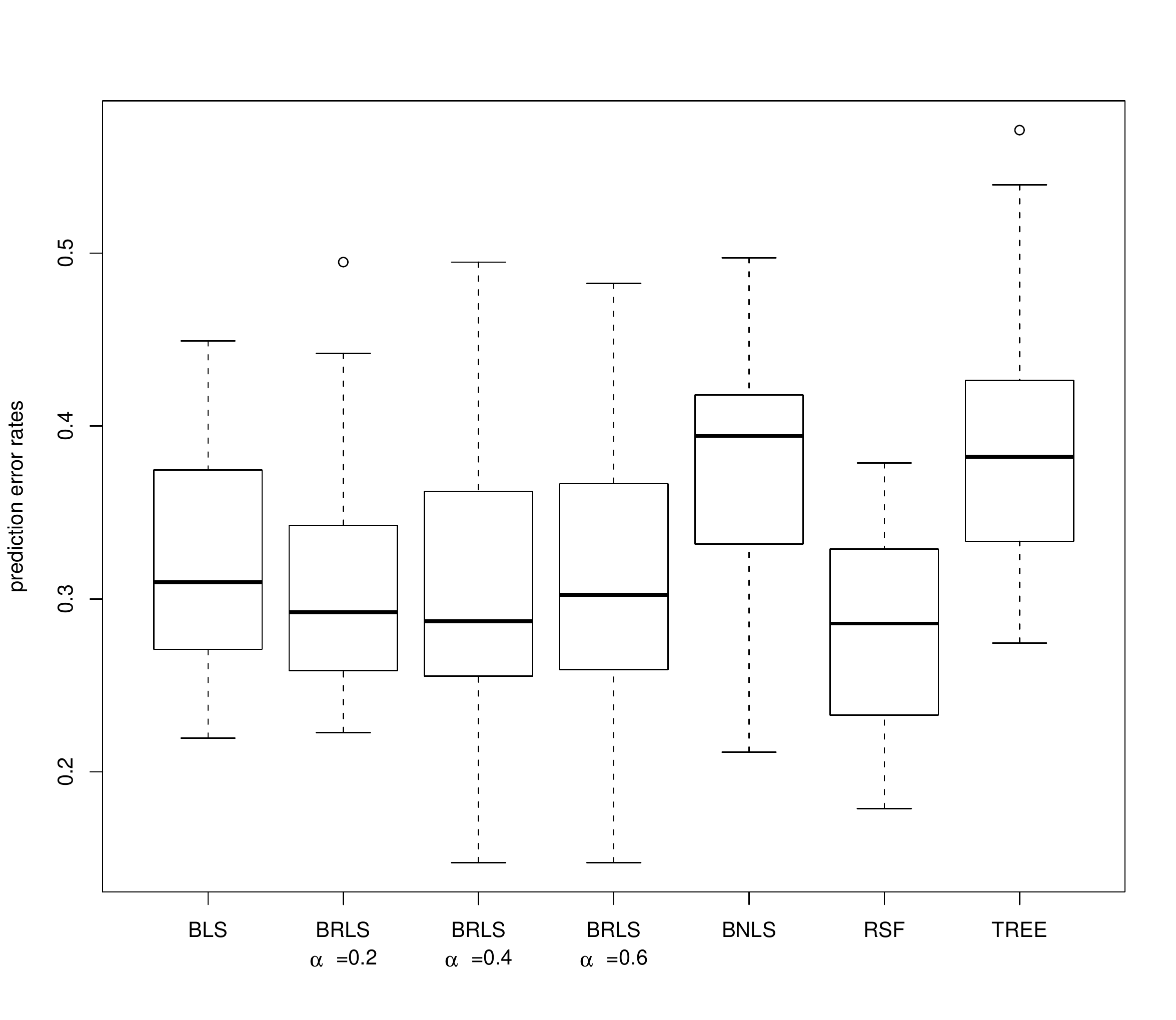}}
\caption{Boxplots of the prediction error rates for seven methods : Bootstrap Lasso selection, Bootstrap Randomized Lasso selection (three values of $\alpha$), Bootstrap node-level selection, Random Survival Forest and a simple survival
tree.}
\label{error_breast}
\end{figure}

\begin{table*}
	\centering
		\begin{tabular}{l|ccc}
			\hline 
			 & Mean & Standard deviation & Median \\
			\hline
			BLS & 0.319 & 0.066  & 0.310   \\
			BRLS $\alpha = 0.2$ & 0.309  & 0.075  & 0.292 \\
			BRLS $\alpha = 0.4$ & 0.309  &  0.081 & 0.287 \\
			BRLS $\alpha = 0.6$ & 0.311  & 0.086 & 0.302 \\
			BNLS & 0.376  & 0.073 & 0.394  \\
			RSF & 0.279 & 0.062 & 0.286  \\
			TREE & 0.389 & 0.0741  & 0.382\\ 
			\hline
		\end{tabular}
		\caption{Mean, standard deviation and median of error rates for the different procedures:
 Bootstrap Lasso selection, Bootstrap Randomized Lasso selection (three values of $\alpha$), Bootstrap node-level selection, Random Survival Forest and a single survival
tree.}
\label{moy_breast}
\end{table*}

\subsubsection{Variables selected in the final model}

Figure $3$ shows the covariates selected by the Cox based
procedures BLS and BRLS according to the choice of the inclusion
frequency $\kappa$. Only 23 covariates of the 75 initial covariates
appear in the selection done by the procedures applied to the boostrapped
samples (i.e.  with $\kappa=0$) and no clinical covariate is selected.  For $\kappa = 0.2$, six covariates
are selected by BLS whereas only four are selected by the BRLS
procedures whatever the value of $\alpha$ (these four covariates are a
subset of the previous six).  We notice a clear gap between the four
first selected variables and the other ones in the BRLS procedures. We
can observe that the BLS and BRLS procedures find the same four most
relevant covariates but not in the same order: \texttt{PRC1},
\texttt{ZNF533}, \texttt{QSCN6L1} and \texttt{IGFBP5.1}.

\begin{figure}
	\centering
		\begin{tabular}{cccc}
			\resizebox{8cm}{!}{\includegraphics{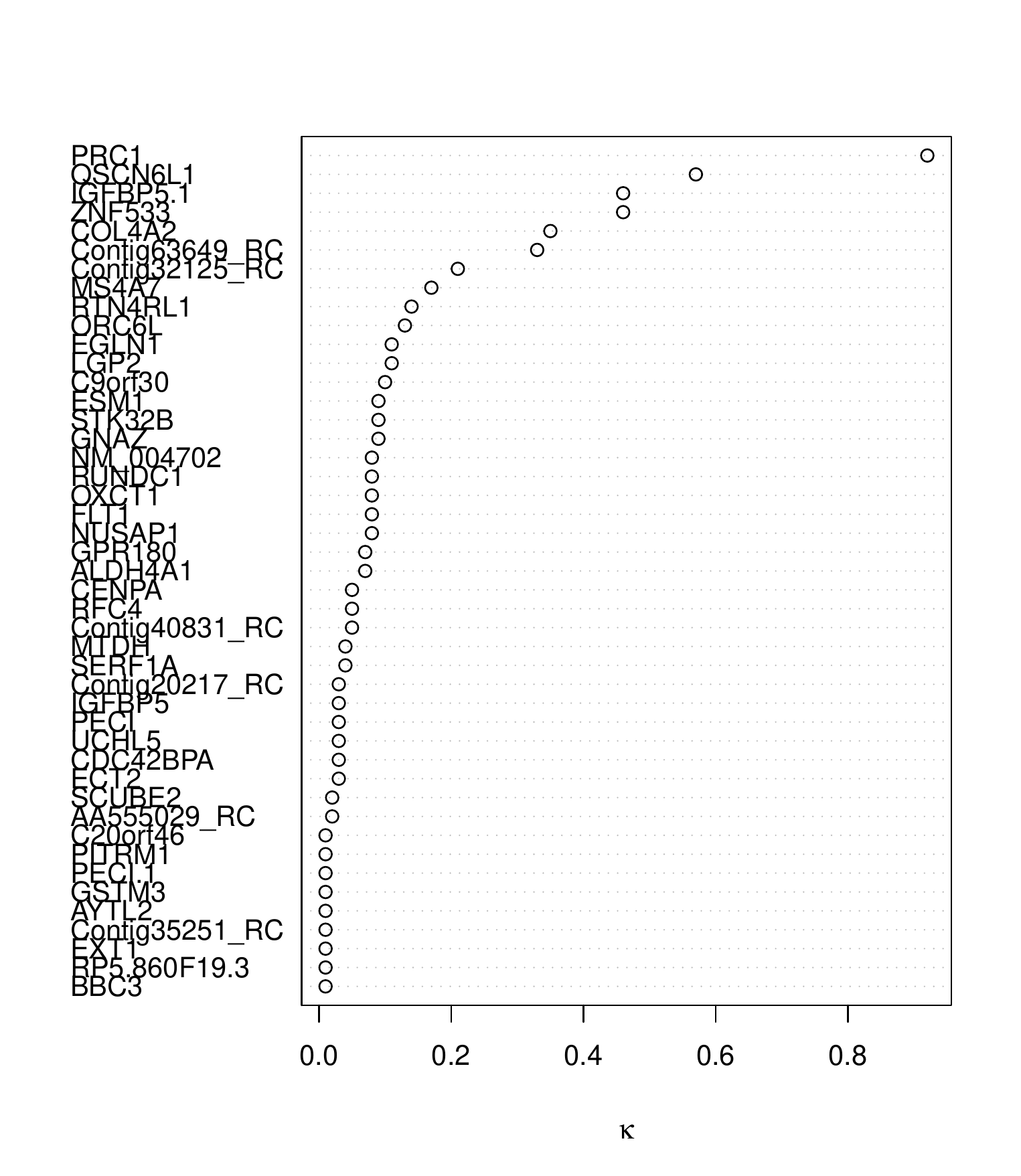}} & \resizebox{8cm}{!}{\includegraphics{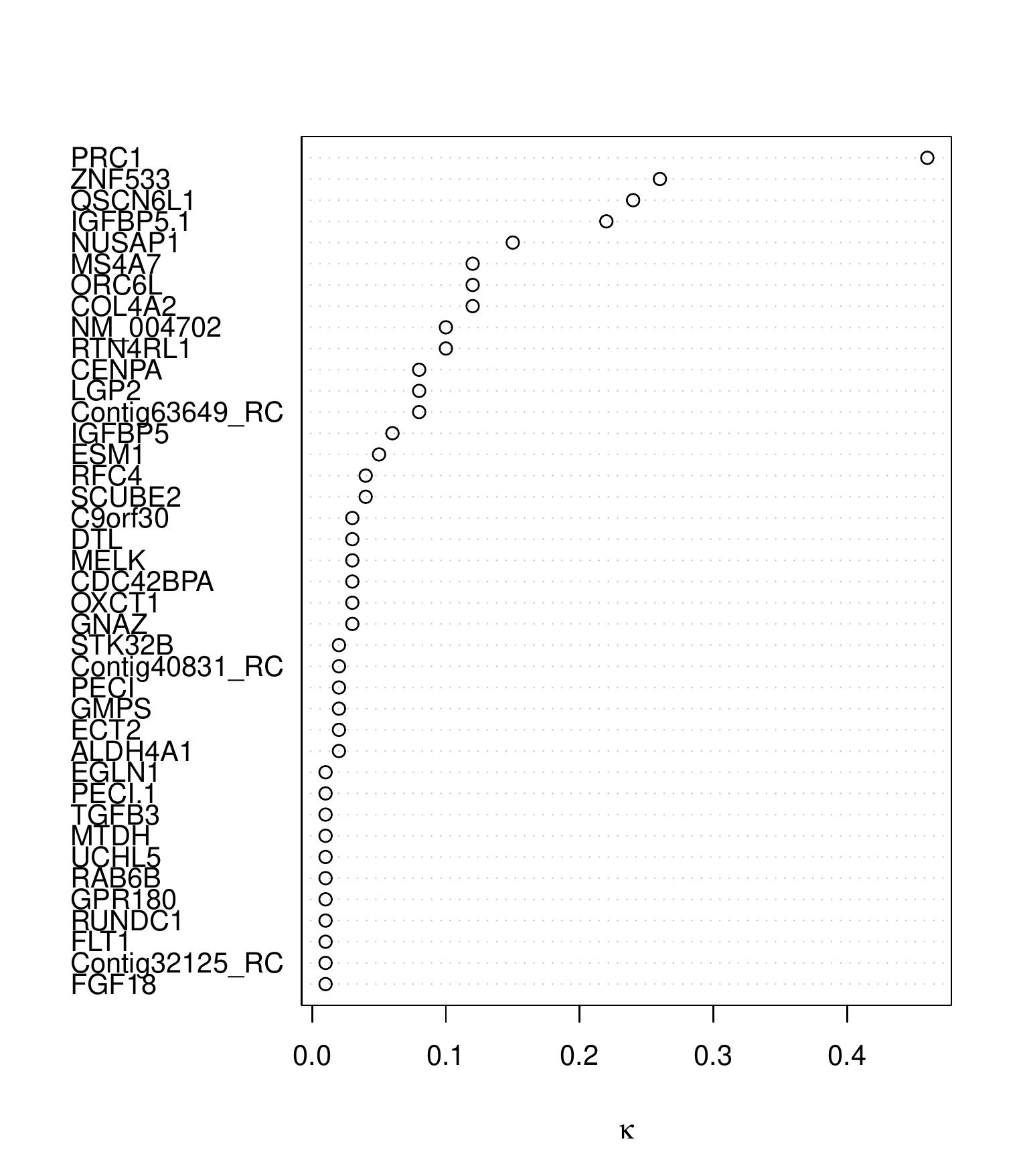}}\\
			\textit{1. BLS, $\lambda = 0.4$} & \textit{2. BRLS, $\alpha = 0.2$, $\lambda = 0.4$} \\
			\resizebox{8cm}{!}{\includegraphics{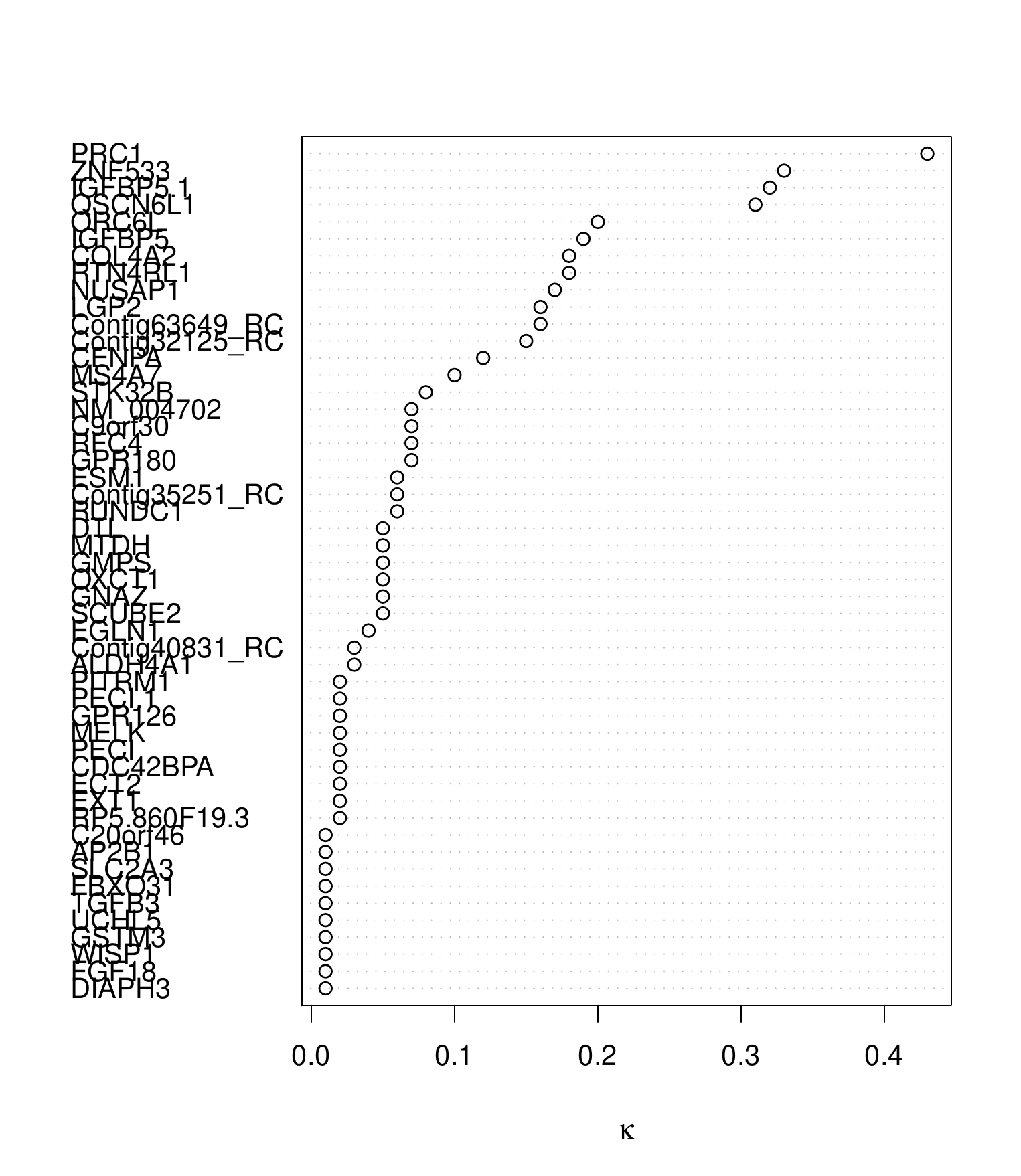}} & \resizebox{8cm}{!}{\includegraphics{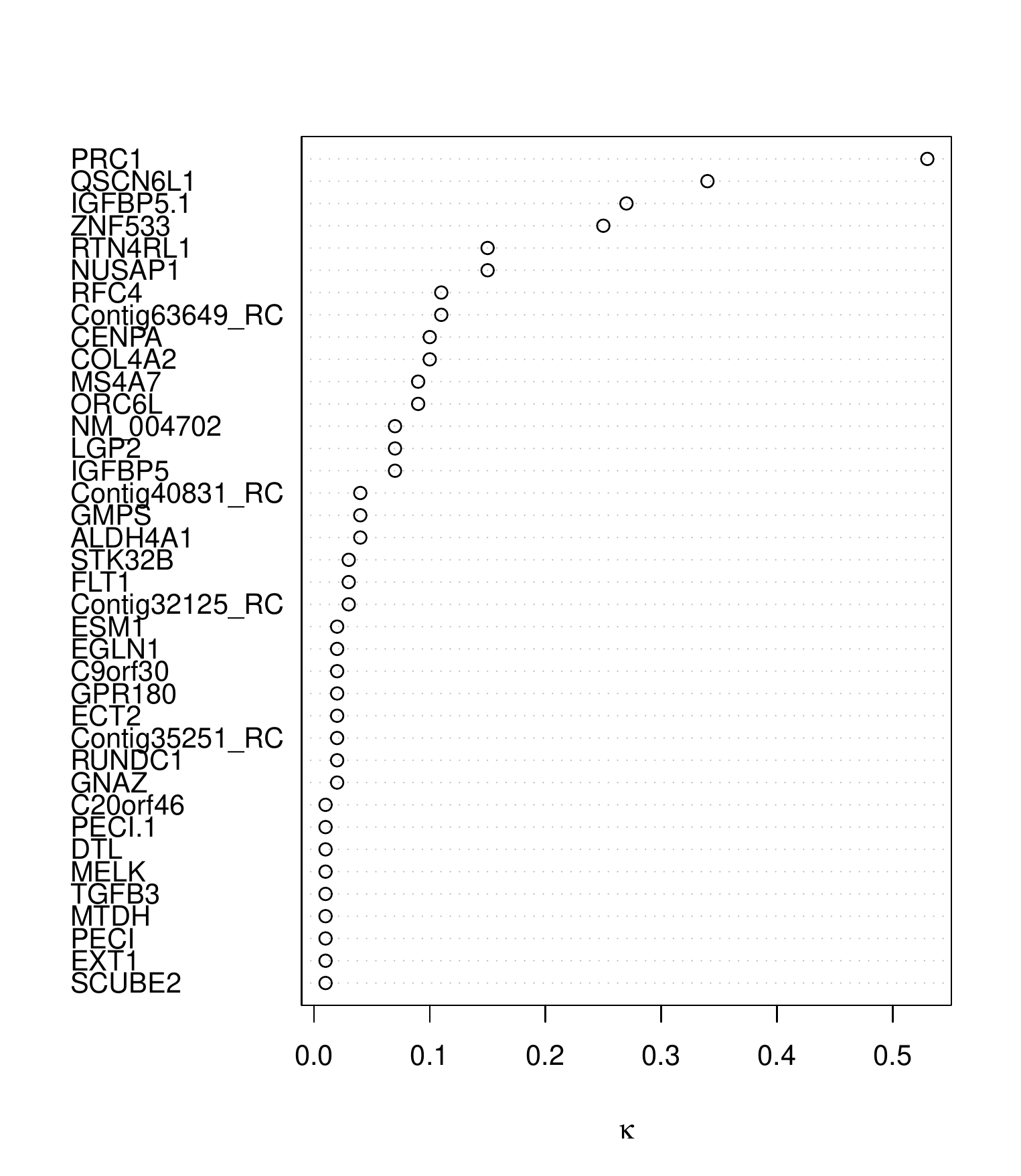}} \\
			\textit{3. BRLS, $\alpha = 0.4$, $\lambda = 0.4$} & \textit{4. BRLS, $\alpha = 0.6$, $\lambda = 0.2$} \\
		\end{tabular}
	\caption{Selected covariates with respect to the inclusion frequency $\kappa$ for Bootstrap Lasso Selection and Bootstrap Randomized Lasso Selection}
	\label{cox_breast}
\end{figure}

For the tree-based procedures  RSF and BNLS, figure $4$ and $5$ show that the same most discriminant variable is \texttt{ZNF533}.  BNLS selects three other covariates among the first most important variables selected by RSF: \texttt{COL4A2}, \texttt{PRC1} and \texttt{N}.
The selection with a single survival tree procedure exhibits the same first discriminant variable \texttt{ZNF533} followed by \texttt{PRC1}, \texttt{RP5.860F19.3}, \texttt{HRASLS}, \texttt{IGFBP5}, and \texttt{SCUBE2}. But the additional selected covariates by the CART algorithm  never appear in the stable selection procedures. We can also notice that no clinical factor is included in the survival tree built by CART procedure  unlike the BNLS and RSF procedures.

As far as the comparison of the Cox based methods and survival trees
methods is concerned, we notice similarity between the selected
covariates: for example, \texttt{ZNF533} is found in all
models. Regarding clinical factors, none was selected in Cox models
contrary to the survival trees based procedures.  As they can take
into account interactions between covariates, tree based procedures
show that \texttt{N}, the number of affected lymph nodes and the age
of the patient at diagnostic are relevant predictors in
metastasis-free survival. Notice that these two factors are also known
to be clinically relevant.

\begin{figure}
	\centering
			\vspace{-1.5cm}\resizebox{17cm}{!}{\includegraphics{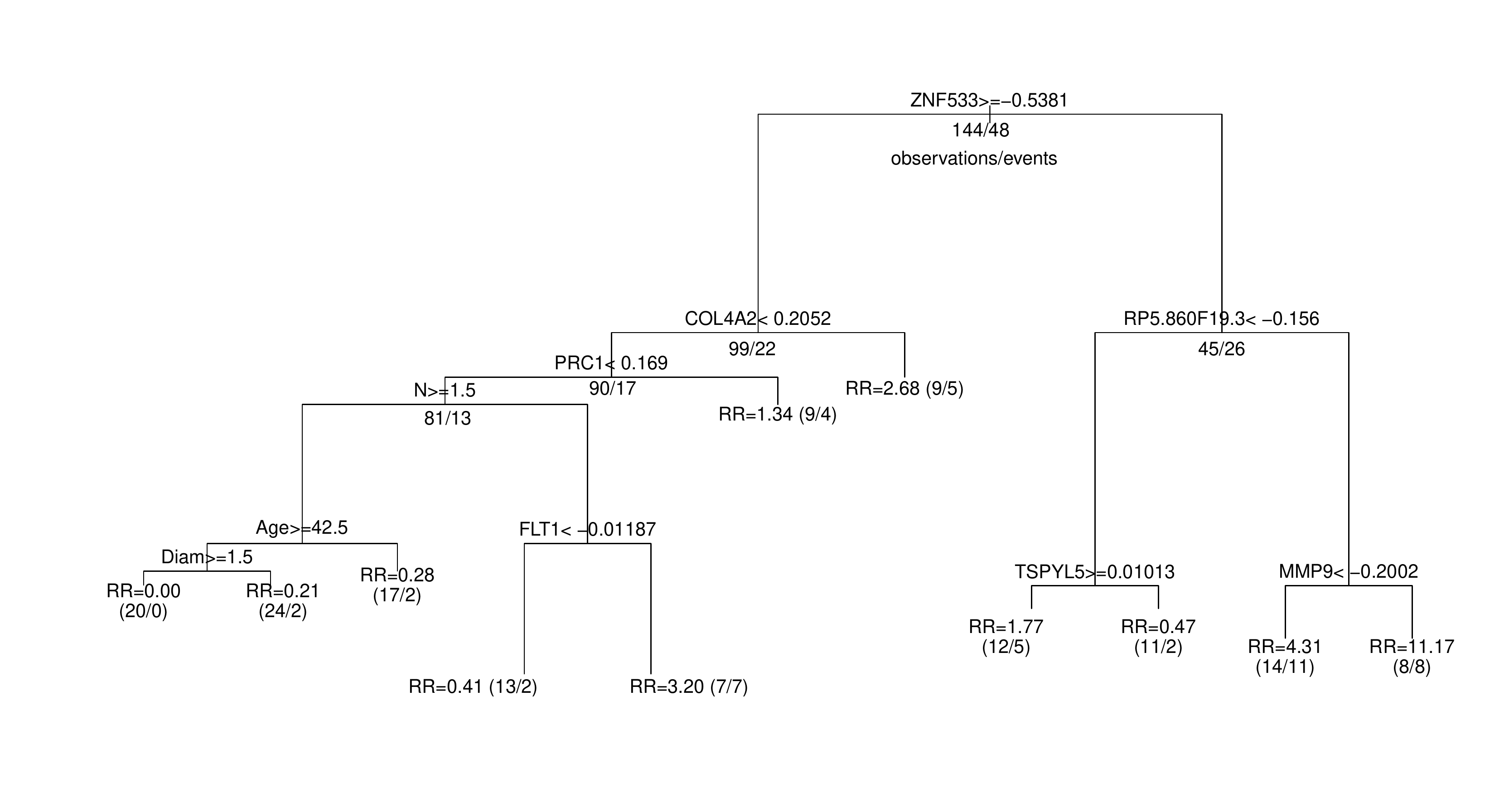}}
	\caption{The final tree obtained by the Bootstrap Node-Level Selection.}
	\label{bnl_breast}
\end{figure}

\begin{figure}
	\centering
		\vspace{-2.5cm}\resizebox{8cm}{!}{\includegraphics{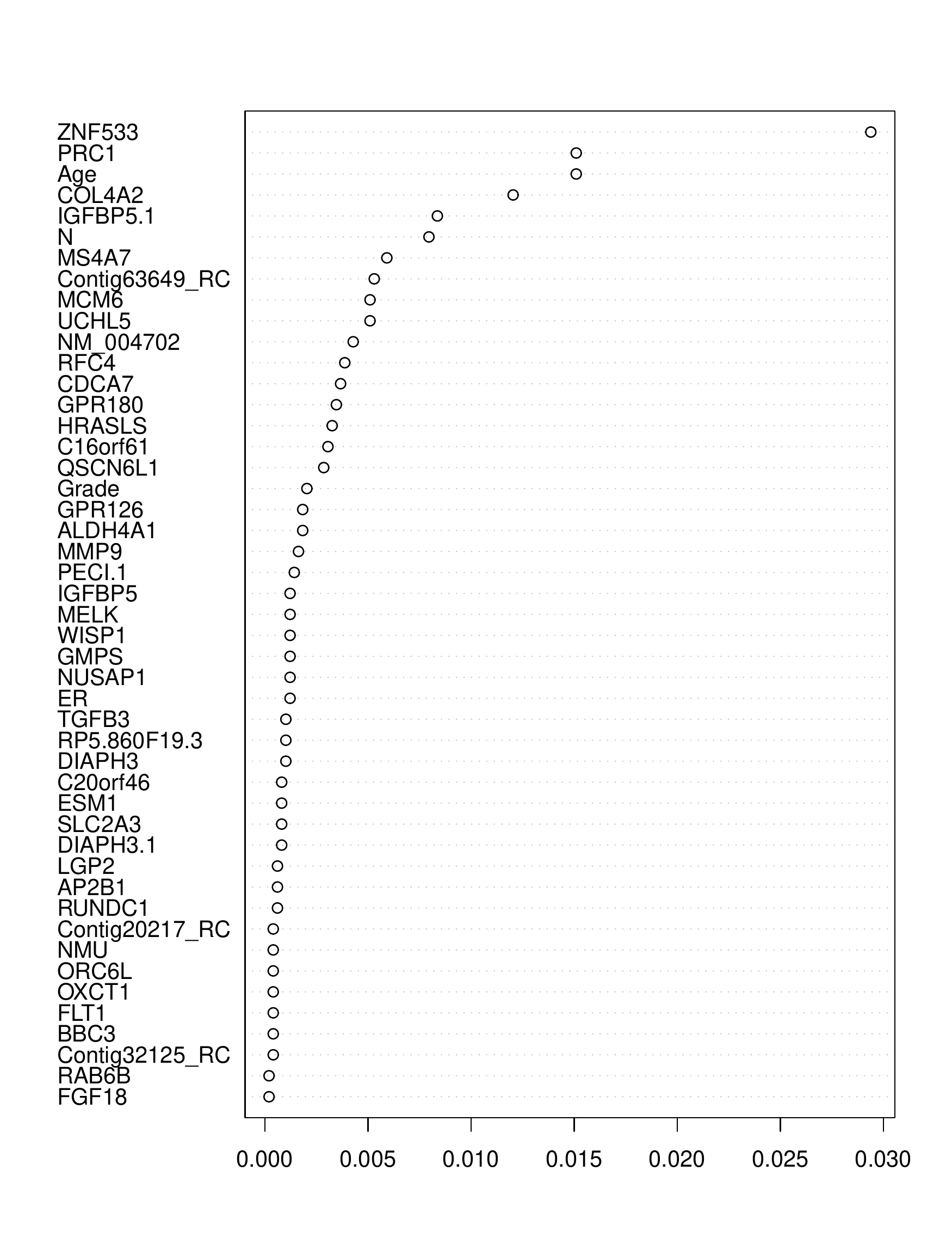}}\\
	\caption{Importance of variables for Random Survival Forest.}
	\label{rsf_breast}
\end{figure}

\subsection{Results for the fertility data set}

\subsubsection{Description}
The fertility data were obtained from 2138 couples consulting for male infertility during the period from 2000 to 2004 at Toulouse Male
Sterility Center (TMSC) located in Universitary Public Hospital (France). Patients were followed from entry and during treatment by an
 andrologist specialist until either discontinuation of treatment or delivery of an alive infant. The maximum follow-up duration is 9 years.
The outcome assessment was based on the delivery of an alive infant obtained at TMSC (pregnancies after
medical treatment - medecine and/or surgical treatment, or assisted reproductive technologies (ART) - as well as spontaneous
pregnancies). The event considered here is the birth of an alive infant  and right-censored events correspond to miscarriage or loss to follow-up. The ``survival'' time is the delay in months from the first visit of the couple to the birth of its alive infant. 
We will work on the subset of the $1773$ couples with covariates  without missing values. 40\% of the couples succeed in their parental project, leading to a censoring rate of 60 \%.
In agreement with clinicians, we decided to keep 32 covariates, among which:
\begin{itemize}
	\item \textbf{for the man}: age, clinical investigation including medical histories as: histories of orchitis, epididymitis, urogenital infections (UI), sexually transmitted infection (STI), inguinal hernia, testis trauma, cryptorchidism (testicular maldescent), cancer, as well as testis, epididymis or vas deferens surgery, and clinical examination as: aspects of scrotum, testicular migration (presence of the testes into the scrotum and their position), epididymides and vasa deferentia evaluation, presence of a varicocele or hydrocele, testicular volumes, and if he received a non-ART treatment (both medical and surgical, including pharmacological and hormonal treatment, varicocelectomy, and vas deferens/epididymis surgery). 
	\item \textbf{for the  woman}: age, tubal factor, ovarian factor, cervical factor, ovulation, and if she received a non-ART treatment (hormonal treatment).
	\item \textbf{for the couple}: fecundity type (primary or secondary), infertility duration, type of ART including IUI (intra-uterine insemination), IVF (in vitro fertilization), ICSI (intracytoplasmic sperm injection) using male sperm cells, and ART with donor sperm.
\end{itemize}

\subsubsection{Prediction error rates}

As for the breast cancer data set, we chose $\lambda$ and $\kappa$ graphically. A first sight of figure $6$  reveals  that the contribution of the BRLS procedure with regard to the BLS procedure for the choice of the cut-off values is not obvious. For BLS and BRLS, only two variables seem to be the most relevant factors  whatever the value of $\lambda$. But for BLS and BLRS with $\alpha = 0.6$, five other covariates follow the two first when $\lambda = 0.4$, for a value of $\kappa=0.3$ whereas for BRLS with $\alpha=0.2$ and $\alpha=0.4$, no group of variables seems relevant. So we decided to choose a fixed penalty $\lambda=0.4$ for a value of $\kappa=0.3$ for BLS and BRLS.

\begin{figure}
	\centering
		\begin{tabular}{cccc}
			\resizebox{8cm}{!}{\includegraphics{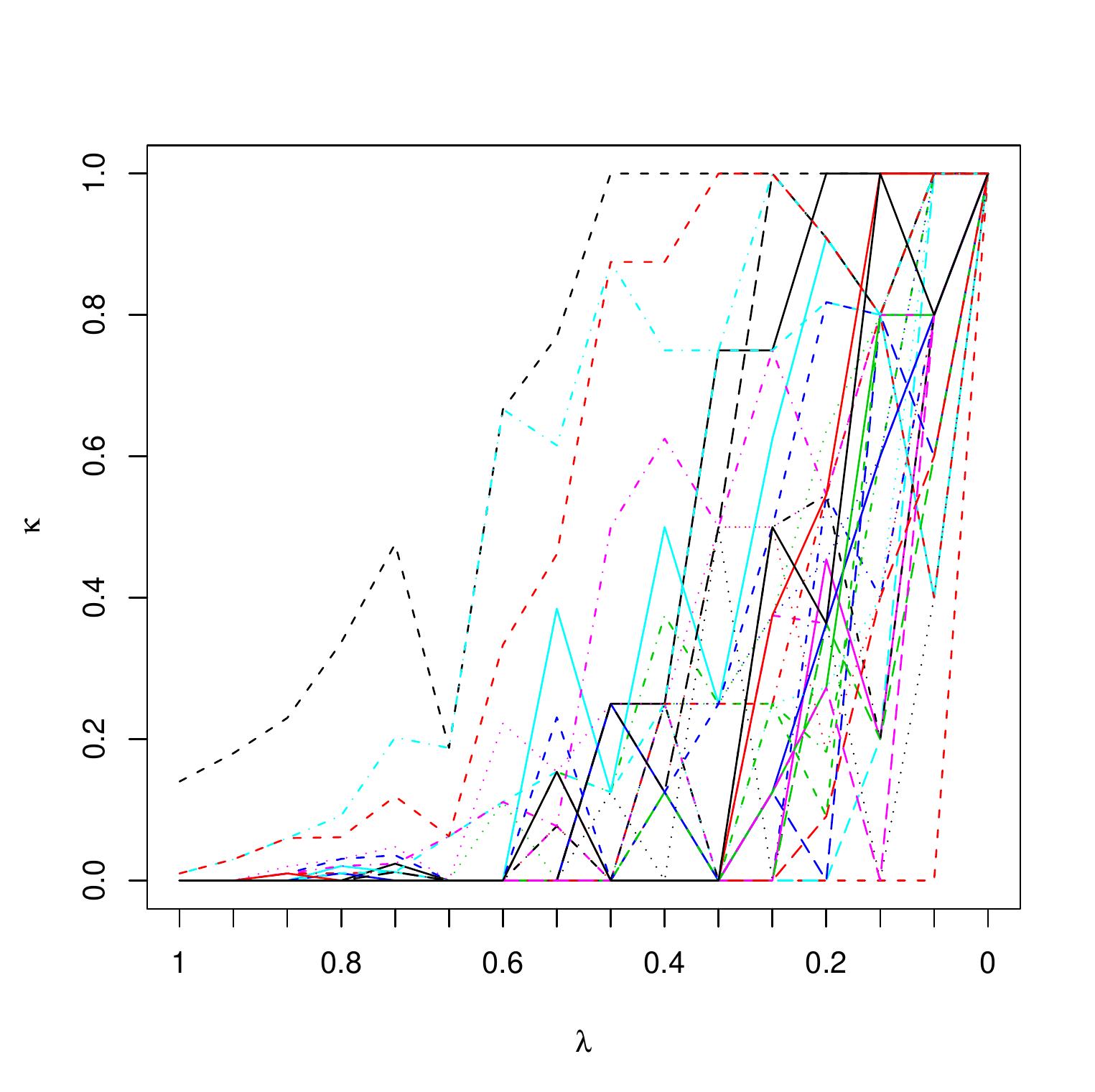}} & \resizebox{8cm}{!}{\includegraphics{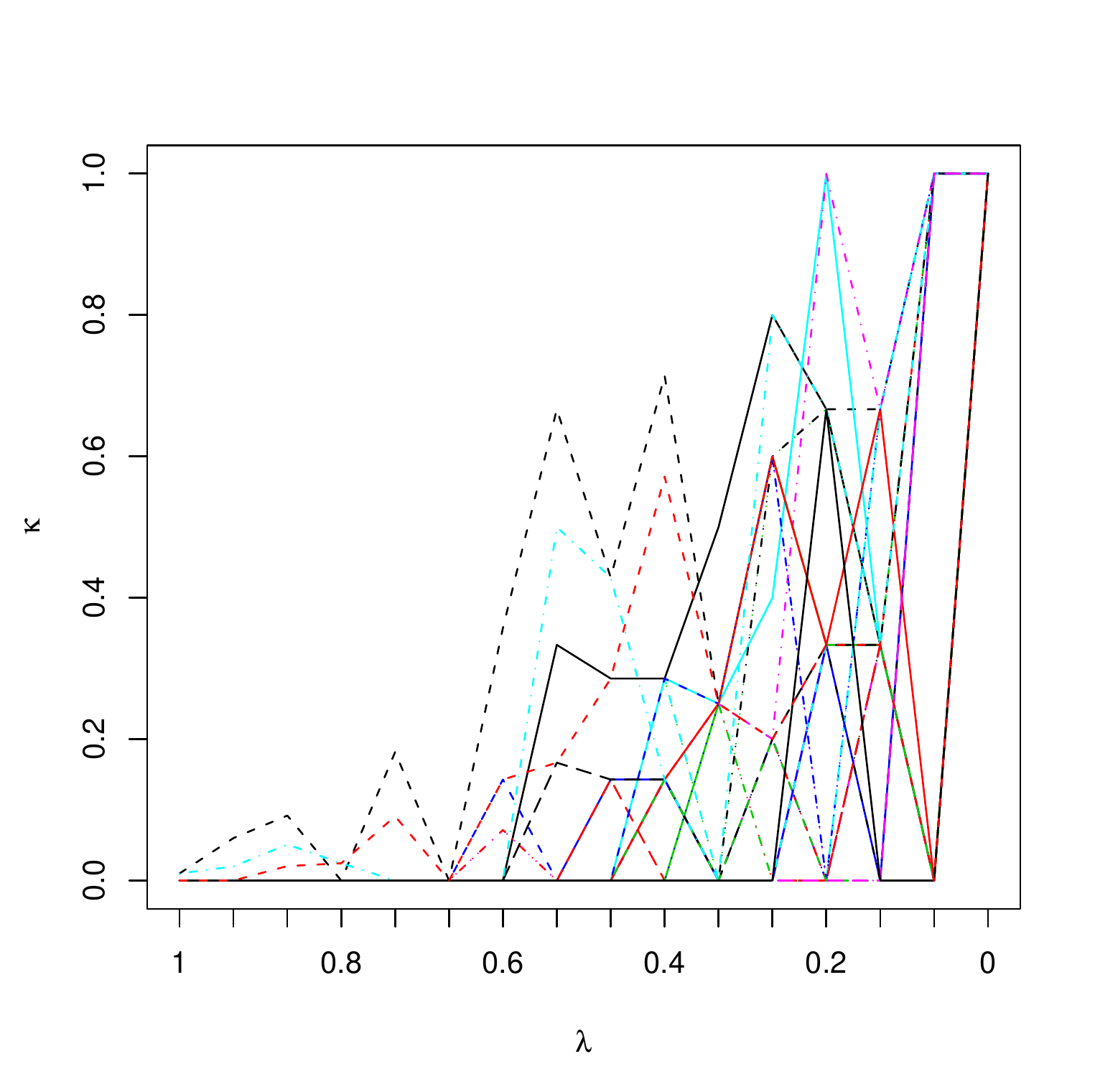}}\\
			\textit{1. BLS} & \textit{2. BRLS, $\alpha = 0.2$} \\
			\resizebox{8cm}{!}{\includegraphics{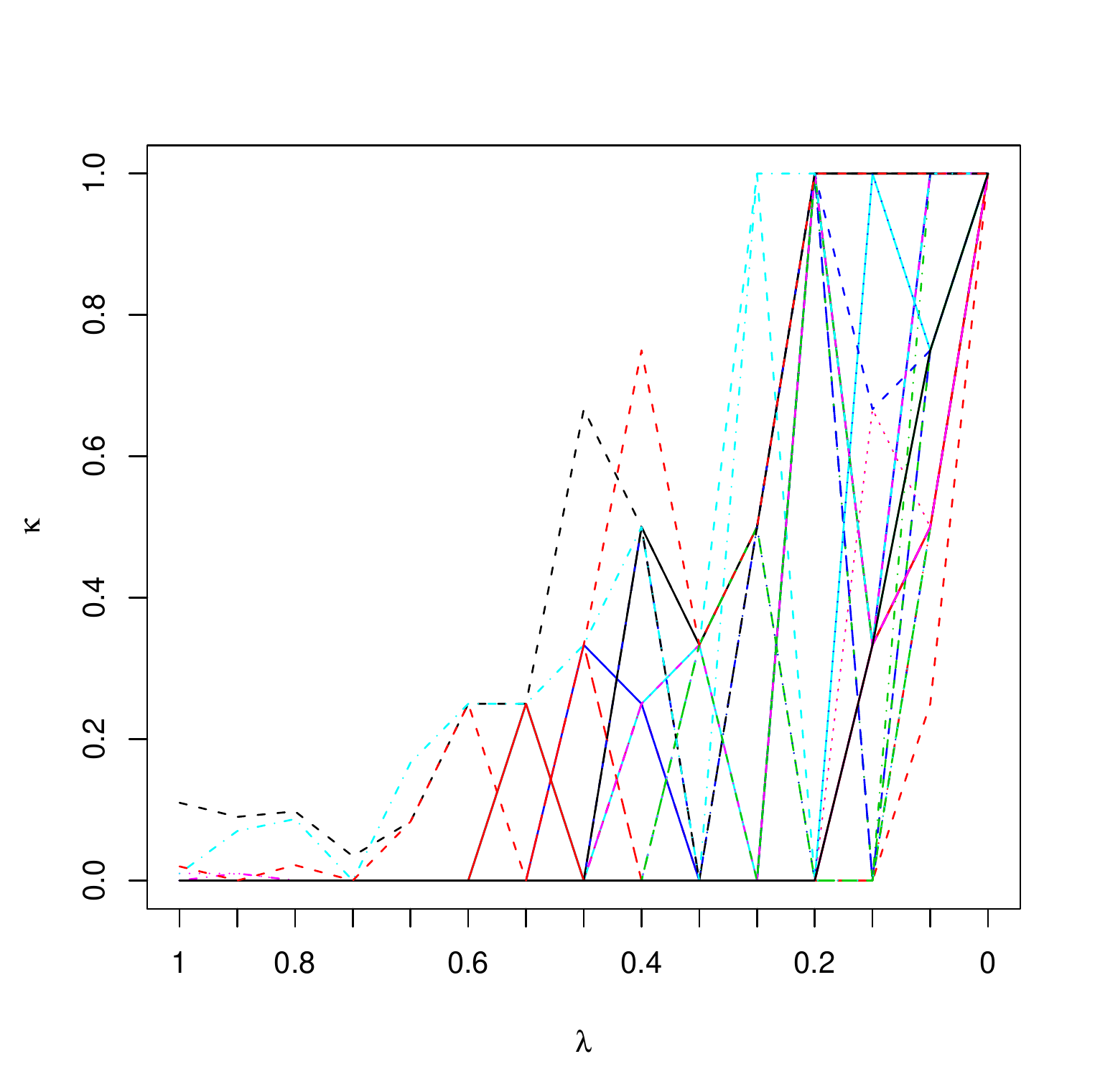}} & \resizebox{8cm}{!}{\includegraphics{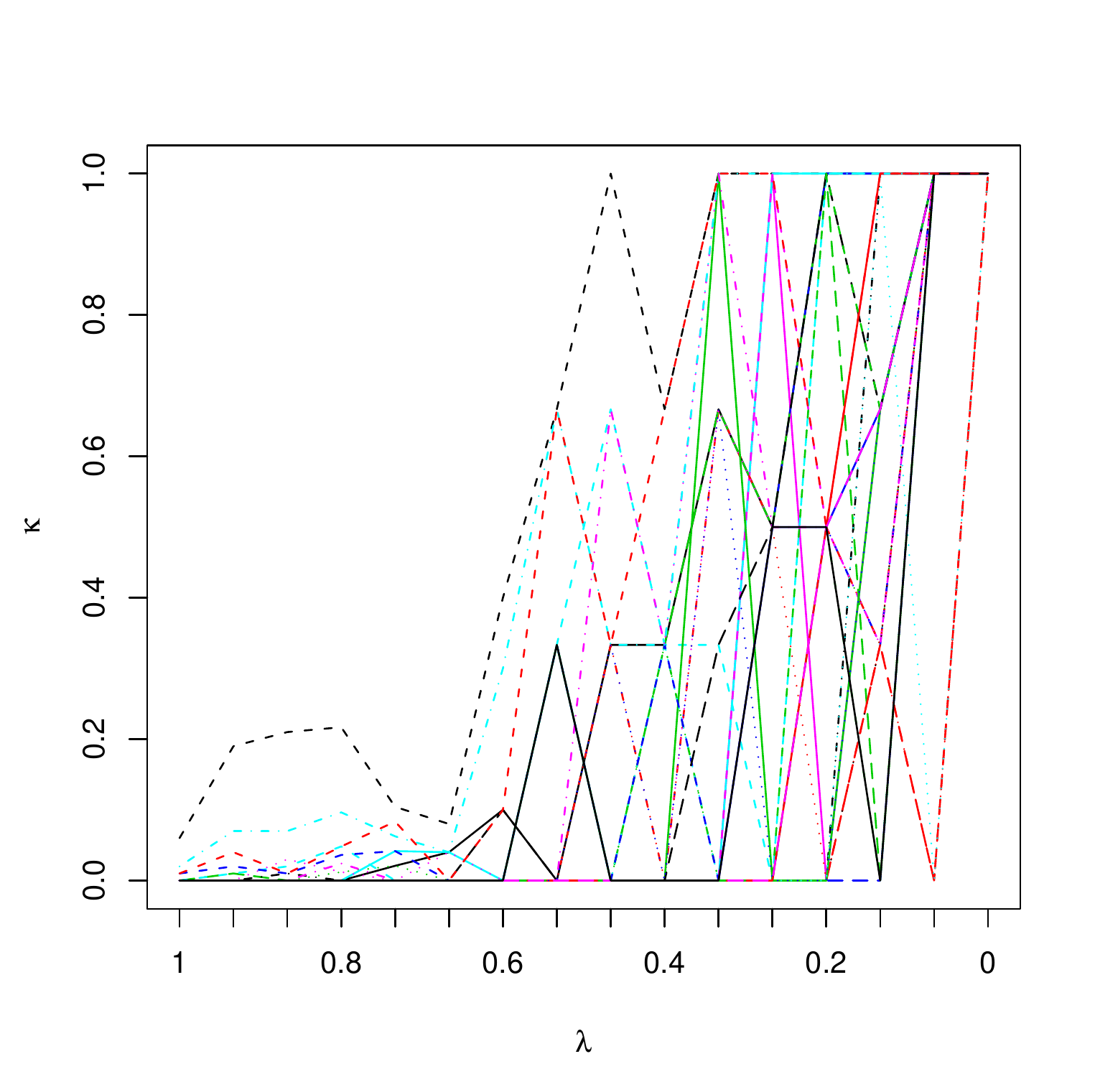}} \\
			\textit{3. BRLS, $\alpha = 0.4$} & \textit{4. BRLS, $\alpha = 0.6$} \\
		\end{tabular}
	\caption{Inclusion frequencies $\kappa$ with respect  to $\lambda$ for Bootstrap Lasso Selection and Bootstrap Randomized Lasso Selection.}
	\label{lasso_ferti}
\end{figure}

For the BSS procedure, we decide to choose a value of $\kappa = 0.5$ which seems to split the set of covariates in two distinct subgroups as suggested by figure $8$. Thus, the first eight  variables seem to be the most relevant to predict the duration to the birth. For the BNLS procedure, we observe by a 10-fold cross-validation procedure that the optimal value of the complexity parameter is $cp=0.0035$.

The boxplots of the prediction error rates for the five procedures are presented in figure $7$ and table \ref{moy_ferti}  shows the summary statistics. Notice  that the mean and median values obtained are higher (between $0.41$ and $0.45$) than those obtained from the data set on breast cancer which reflects the difficulty to predict the delay to the birth. However, the variabilities of the error rates are lower. As seen previously, the RSF method gives the best predictive model. However, it appears that RSF is similar in dispersion with other procedures. Moreover, we can notice that the BNLS procedure is not better than a single survival tree and the same remark can be done between the BSS procedure compared to a single stepwise Cox regression. These results can be explained by the size of the sample, which is sufficient to produce low error rates without bootstrapping.  Regarding BLS and BRLS, these procedures show less variation in error rates but their means and medians  are close to $0.5$ which suggests that these models do no better  prediction than random guessing.

\begin{figure}
\centering
\vspace{-3cm}
\resizebox{12cm}{!}{\includegraphics{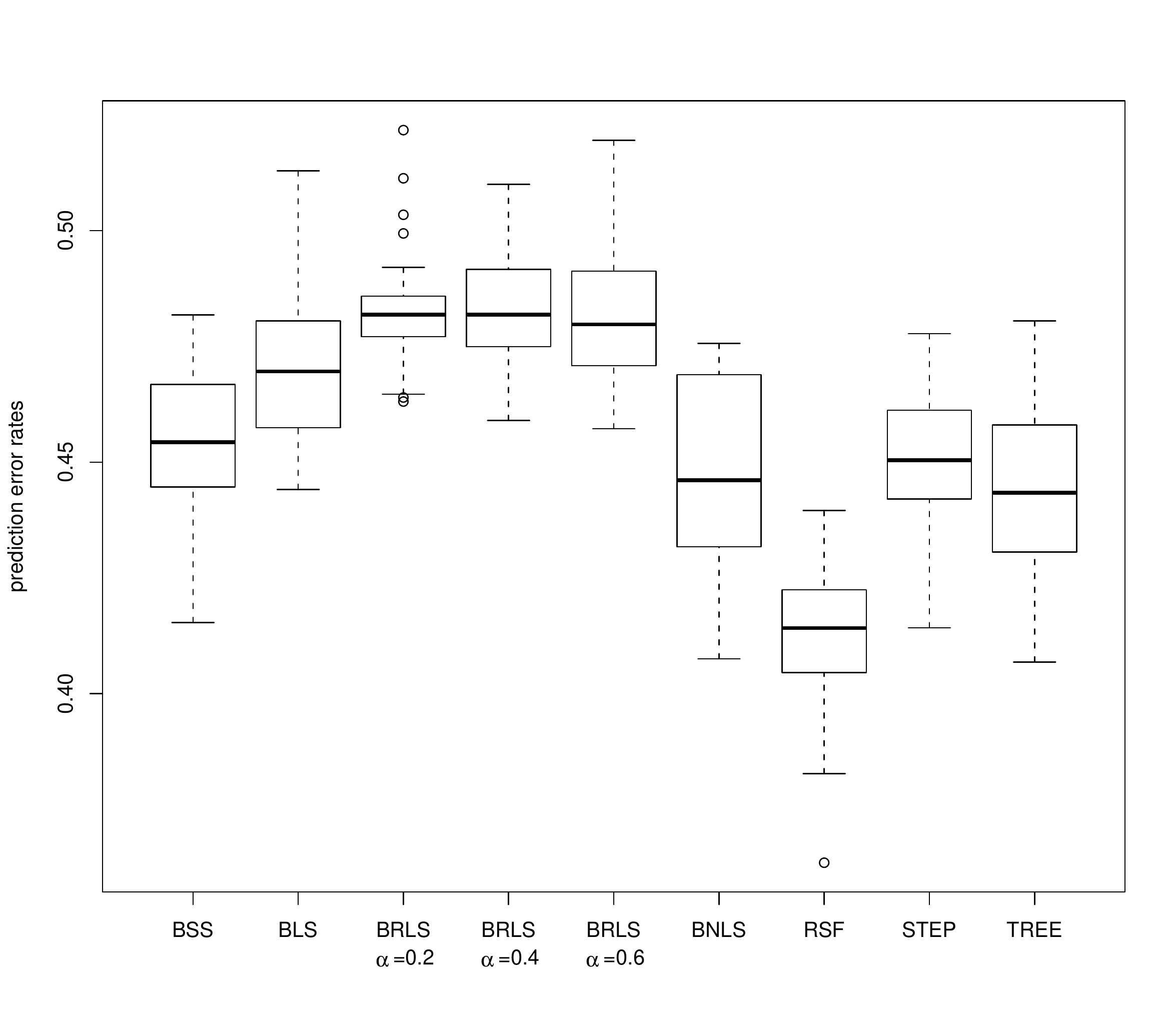}}
\caption{Boxplots of the prediction error rates for nine methods : Bootstrap Stepwise selection, Booststrap Lasso selection, Bootstrap Randomized Lasso selection (three values of $\alpha$), Bootstrap node-level selection, Random Survival Forest, Cox stepwise selection  and a simple survival tree.}
\label{error_ferti}
\end{figure}

\begin{table*}
	\centering
		\begin{tabular}{l|ccc}
			\hline 
			 & Mean & Standard deviation & Median \\
			\hline
			BSS & 0.453  & 0.017  & 0.454   \\
			BLS & 0.471   & 0.017   & 0.470    \\
			BRLS $\alpha = 0.2$ & 0.483   & 0.013  & 0.482 \\
			BRLS $\alpha = 0.4$ & 0.484 & 0.013 & 0.482  \\
			BRLS $\alpha = 0.6$ & 0.482 & 0.015 & 0.480  \\
			BNLS & 0.447 & 0.021 & 0.446  \\
			RSF & 0.413 & 0.017  & 0.414  \\
			STEP & 0.451 & 0.015  & 0.450 \\ 
			TREE & 0.445 & 0.018 & 0.443\\ 
			\hline
		\end{tabular}
		\caption{Mean, standard deviation and median of error
                    rates for the different procedures: Bootstrap Stepwise
                    selection, Booststrap Lasso selection, Bootstrap
                    Randomized Lasso selection (three values of
                    $\alpha$), Bootstrap node-level selection, Random
                    Survival Forest, Cox stepwise selection and a
                    single survival tree.}
\label{moy_ferti}
\end{table*}

\subsubsection{Selected variables in the final model}

If we compare the covariates selected by the different approaches, we
can see in figure $8$ that the first four selected covariates
do not differ for the BLS and BRLS procedures for a value of $\kappa
= 0.3$: we find \texttt{tubal factor}, \texttt{IUI}, \texttt{sperm
  donor} and \texttt{infertility duration}. The additional covariates included by BLS are \texttt{epididymis}, \texttt{varicocelis}, \texttt{inguinal hernia}, \texttt{fecundity type}, \texttt{testicular
  trauma} and \texttt{testicular volume}. For the BSS procedure, for a value of
$\kappa = 0.5$,  we find
in the selected covariates  the first four  covariates selected by BLS but also \texttt{female age},
\texttt{testicular volume}, \texttt{male treatment},
\texttt{varicocelis}, \texttt{testicular trauma}, \texttt{scrotum},
\texttt{female treatment} and \texttt{epididymis}. We observe also that the BSS procedure includes more
variables than the BLS and BRLS procedures, which leads to a lower prediction error
rate.

We can notice that the cut-off value $\kappa=0.3$ appears very clearly
in the graphs of figure $8$ for the BRLS procedures with
$\alpha=0.4$ and $\alpha=0.6$, suggesting that the BRLS procedure is a
good stable variable selection procedure. However, only four
covariates are selected, which leads to poor prediction performances.

If we adjust a single Cox stepwise regression, we find the same
covariates selected by the BSS procedure except for the variable
\texttt{scrotum}. However, \texttt{epididymis}, \texttt{testicular
  trauma} and \texttt{male treatment} are not statistically
significant at the 5\% level in the Cox stepwise model.

\begin{figure}
	\begin{narrow}{-0.5in}{0in}{0in}
		\begin{tabular}{cccc}
			\resizebox{9cm}{!}{\includegraphics{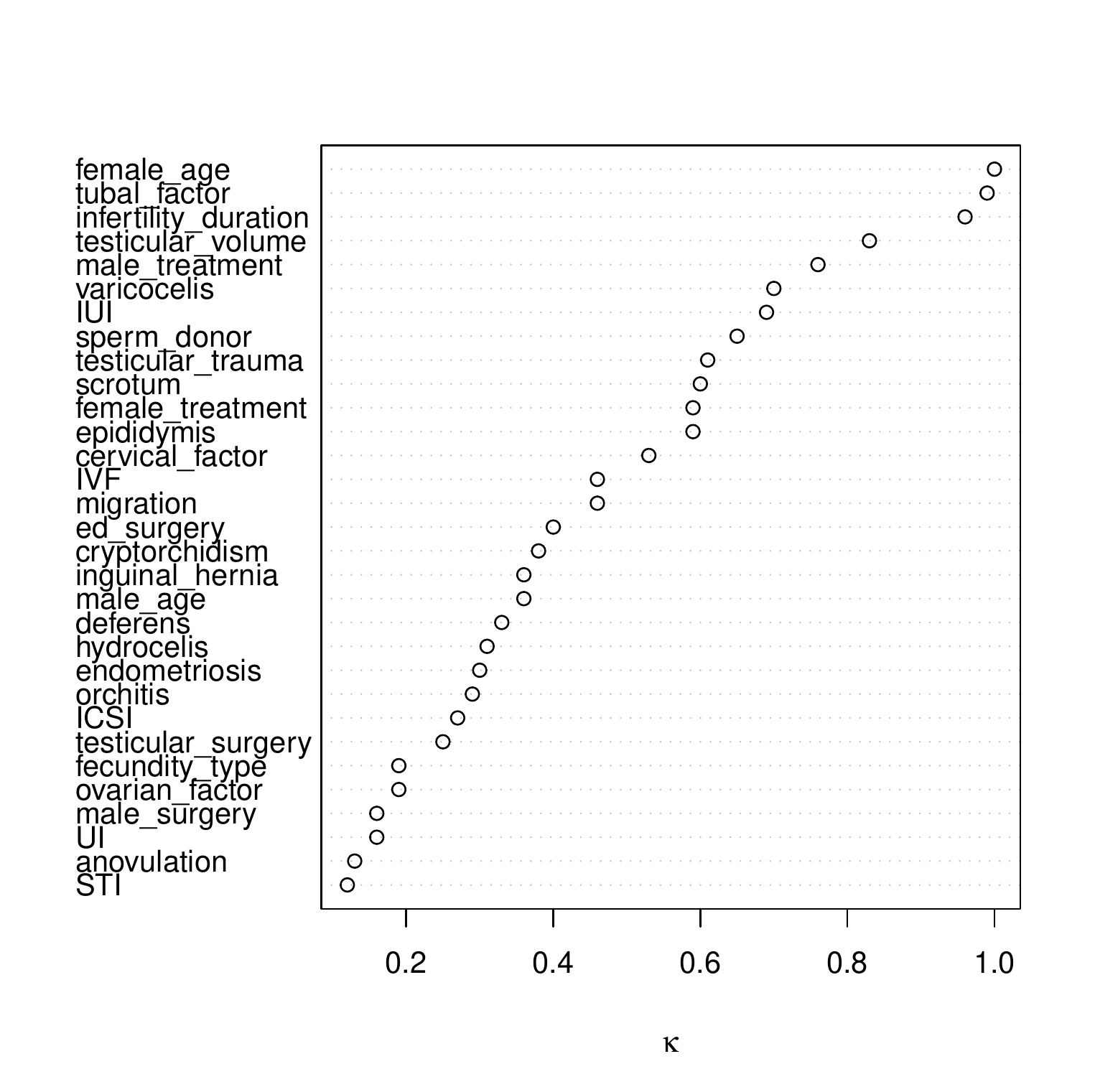}} & \resizebox{9cm}{!}{\includegraphics{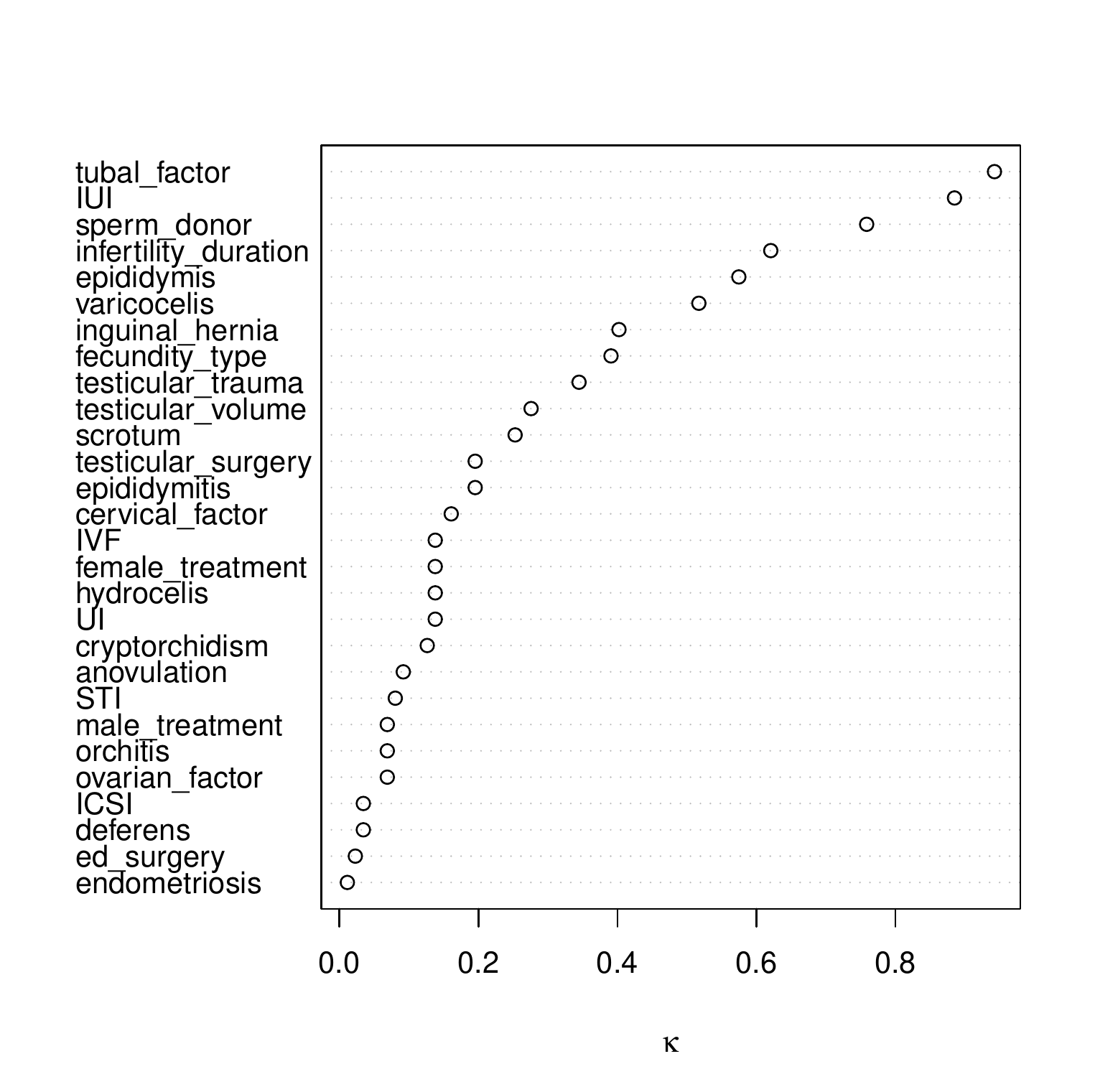}} \\
			\textit{1. BSS} & \textit{2. BLS $\lambda = 0.4$} \\
			\resizebox{9cm}{!}{\includegraphics{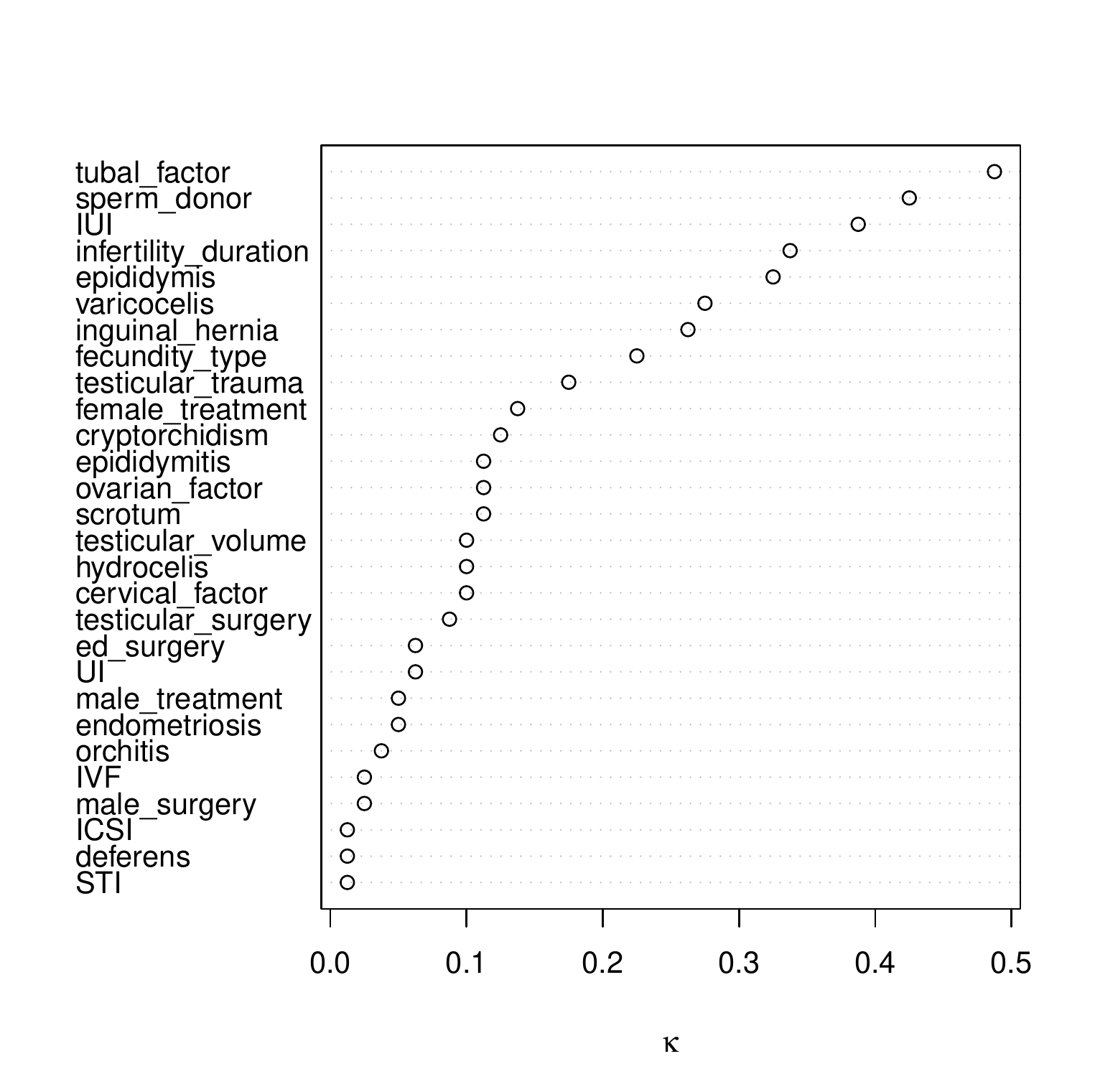}} & \resizebox{9cm}{!}{\includegraphics{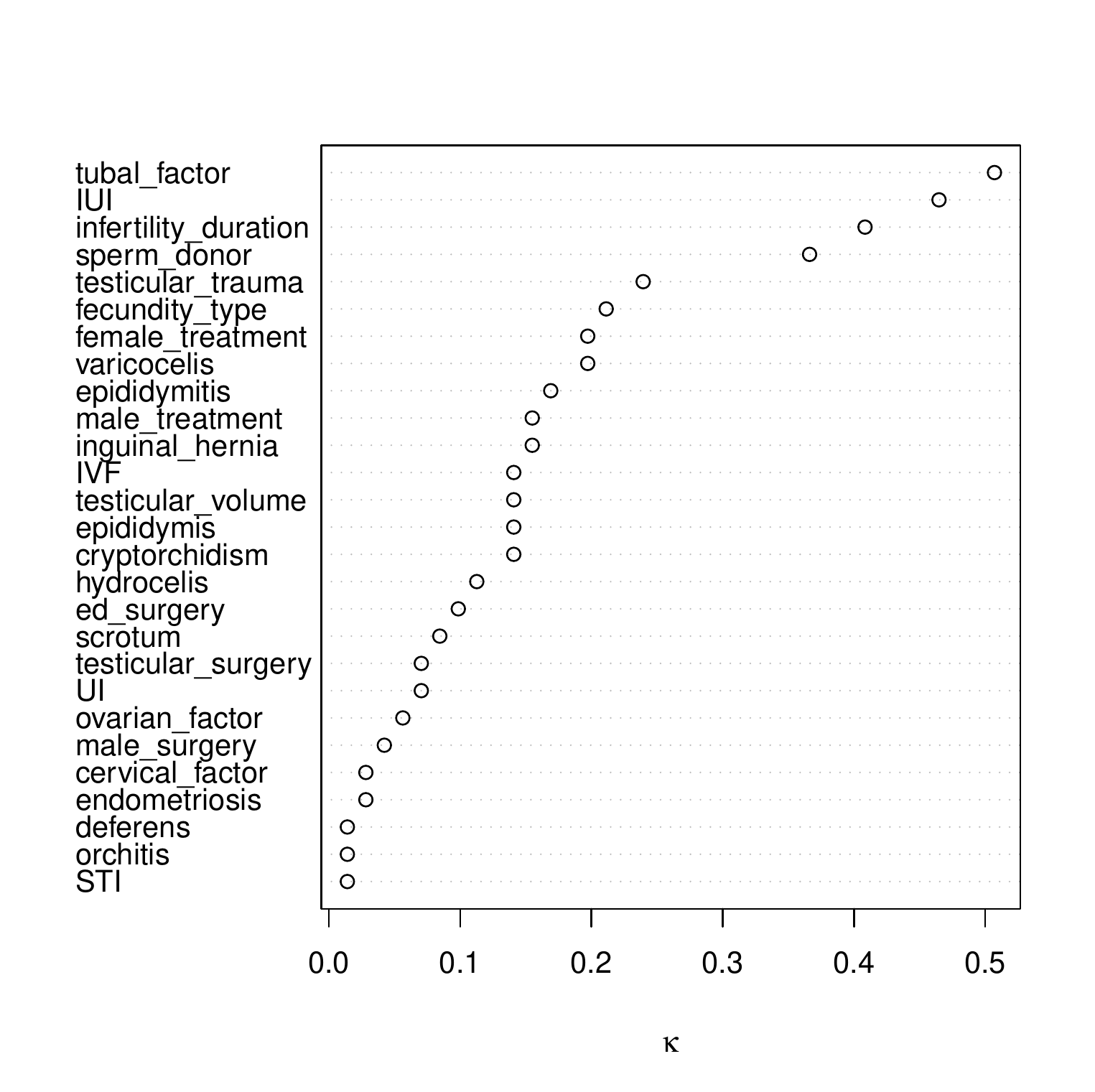}} \\
			\textit{3. BRLS $\alpha = 0.2$, $\lambda = 0.4$} & \textit{4. BRLS $\alpha = 0.4$, $\lambda = 0.4$} \\
		\end{tabular}
	\end{narrow}
	\caption{First part: Selected variables for Bootstrap Stepwise Selection, Bootstrap Lasso Selection, Bootstrap Randomized Lasso Selection and importance of variables for Random Survival Forest.}
	\label{cox_ferti}
\end{figure}
\addtocounter{figure}{-1}
\begin{figure}
	\begin{narrow}{-0.5in}{0in}{0in}
		\begin{tabular}{cccc}
			\resizebox{9cm}{!}{\includegraphics{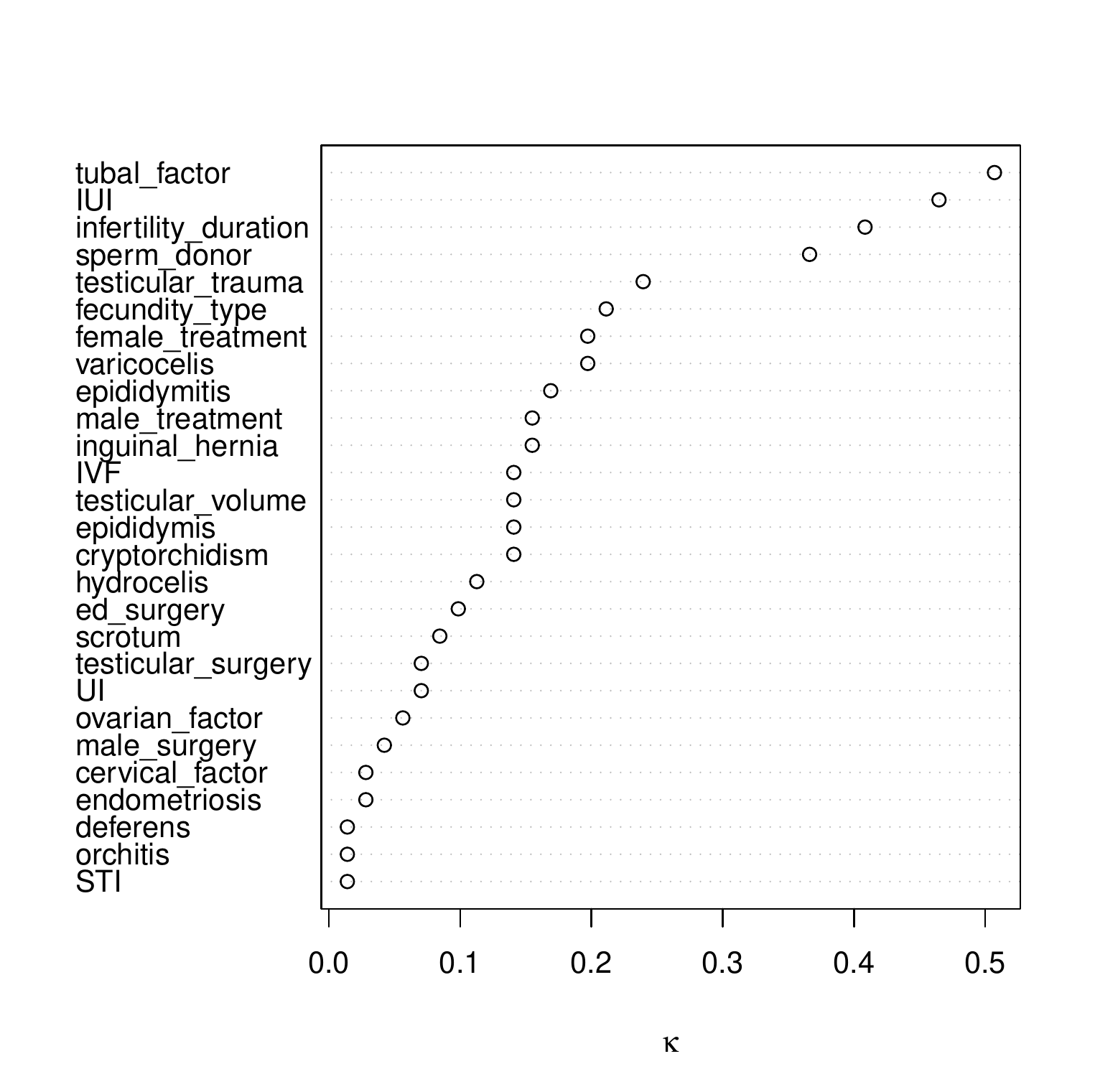}}  & \resizebox{9cm}{!}{\includegraphics{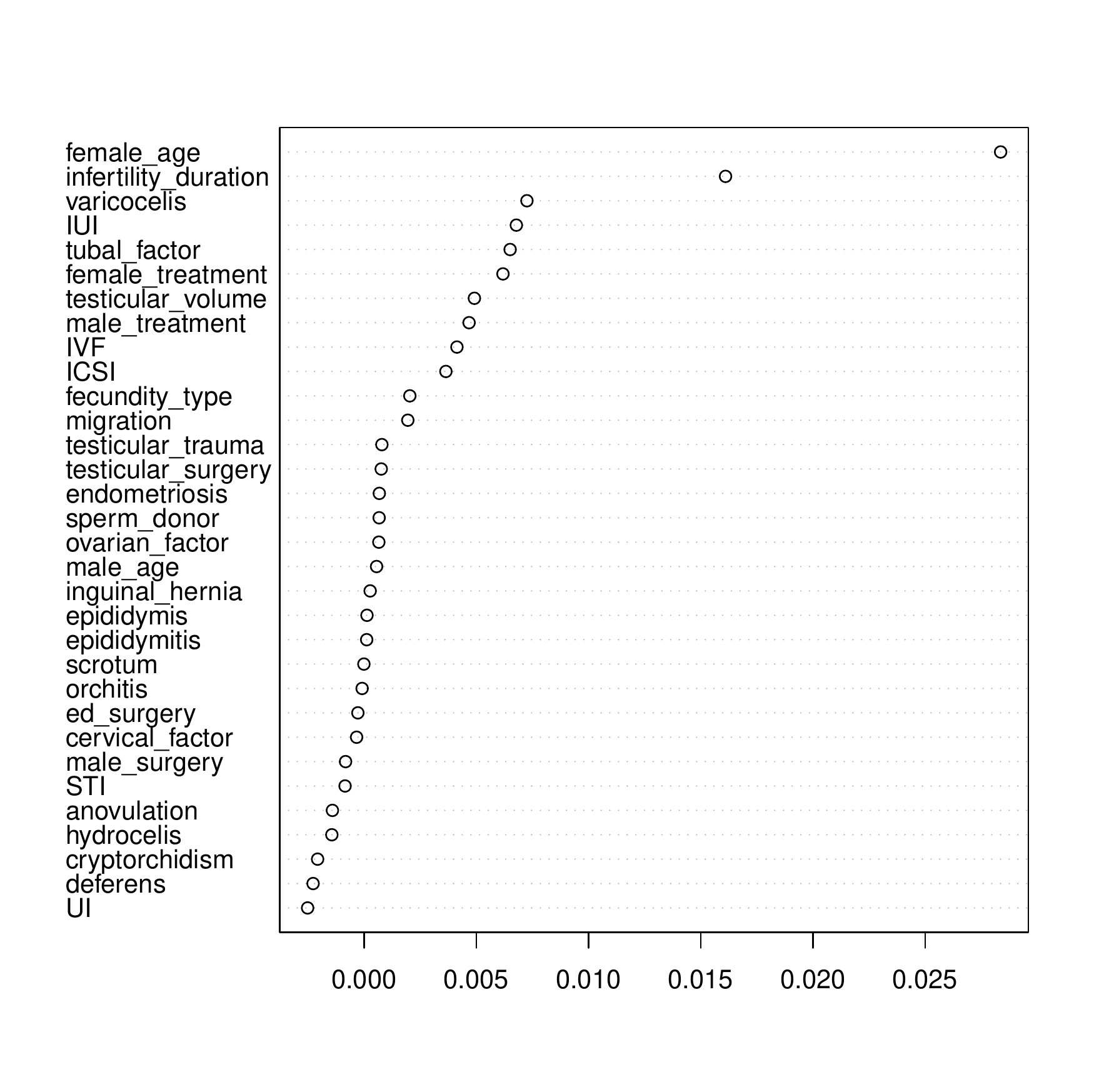}}\\
			\textit{5. BRLS $\alpha = 0.6$, $\lambda = 0.2$} & \textit{6. RSF}\\
		\end{tabular}
	\end{narrow}
	\caption{Second part: Selected variables for Bootstrap Stepwise Selection, Bootstrap Lasso Selection, Bootstrap Randomized Lasso Selection and importance of variables for Random Survival Forest.}
	\label{cox_ferti}
\end{figure}

\begin{figure}
	\centering
			\vspace{-1cm}\resizebox{17cm}{!}{\includegraphics{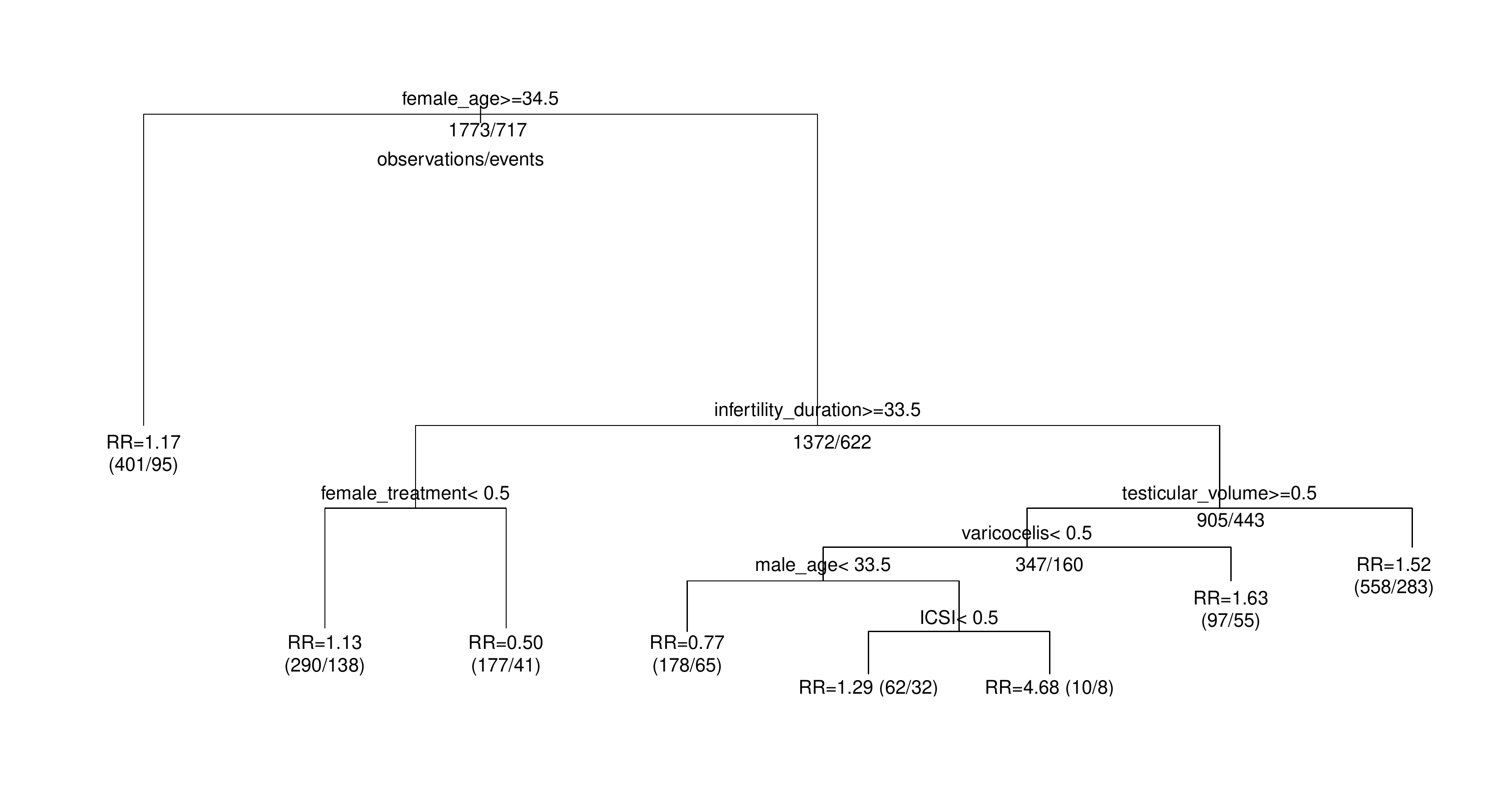}}
	\caption{The final tree obtained from the  Bootstrap Node-Level selection}
	\label{tree_ferti}
\end{figure}

Regarding the tree-based RSF and BNLS procedures, we observe  in figures $8$ and $9$ that the selected covariates are substantially different from the most relevant covariates selected by the Cox based procedures.  The first split, which  corresponds  to  \texttt{female age}, is the same for the two types of procedures but the following splits are different. This may be explained by the fact that the tree based procedures take into account interactions contrary to the Cox model based procedures. Moreover, the covariates selected for the BNLS procedure are found in the most important variables in RSF. The variables selected by a single survival tree are almost the same than those selected by BNLS,  which can be explained by the large sample size of the fertility data set.

\section{Discussion}

\subsection{Computational requirements}

On a 2.4GHz processor, for the fertility data set with 1773 individuals
and 32 covariates, running the RSF procedure is 200 times longer than
running a single survival tree from the CART procedure (which takes
about 1 second), whereas the stepwise algorithm takes about 10
seconds. Compared to the RSF method, the BLS and  BRLS procedures
have similar computing times. The BNLS procedure is much longer than
RSF (4 times) as a bootstrap is realized at each node. Finally, the
BSS procedure, for which a bootstrap sample is used at each stepwise,
is the most expensive procedure in time (8 times longer than RSF). 
Thus, the running times seem reasonable for each procedure taken separately. However, a systematic comparison of errors becomes heavy, as well as the running time to exhibit the optimal complexity parameter $cp$ in the cross-validation procedure for BNLS, and also the penalty $\lambda$ for the BLS and BRLS methods.

\subsection{Prediction error rates}

Compared to the data set on breast cancer, the prediction error rates obtained from the fertility data set are higher, suggesting that data on fertility are more complex and difficult to study and that finding a  good prediction model is a difficult task. Moreover, the sizes of the two data sets are very  different, which could lead to observe discrepancy in the results. However,  we find from the two data sets that the RSF procedure is the best procedure for prediction and that it allows to exhibit the most relevant variables to explain the survival durations. The Cox model and  RSF have been previously compared by Omurlu \textit{et al.} [33] using the Harrell'concordance index. On the basis on  Monte Carlo simulations, they show that the Cox model has the best predictive performance  whatever the size of the sample ($n = 50, 100, 250, 500$). However,  on a real data set on breast cancer, they find that RSF has the lowest prediction error rate. These contradictory results are not surprising if we consider the fact that the simulated data have been generated from a Cox model. 
Moreover, the  RSF procedure  is easy to use and does not require the choice of tuning values as do  the BSS, BLS, BRLS and BNLS procedures. However, even if the selected covariates are identified and sorted by their importance, as no final tree is provided, the RSF results stay a black-box not easy
to interpret and use for clinicians.

We find that the BRLS procedure, whatever the value of $\alpha$, does
not seem to improve the BLS procedure. Although the gap between the
selected variables appears more clearly for BRLS, the final model does
explain these results.  Moreover, the good results obtained with this
method by Meinshausen and Bühlmann [3] may come from the fact that
they presented an ad-hoc example.  As noticed by these authors, we
find on our data sets that choosing a value of $\kappa$ equal to 1, as
suggested by Bach [4], is too restrictive.

Even if some authors [12, 13] showed that the bootstrap method adapted to the
stepwise algorithm improves the stability of the variable selection,
our results suggest no improvement with the BSS procedure. As a matter
of fact, the BSS algorithm does not converge on the breast cancer data
set containing too few events with respect to the number of the
covariates. On the other side, for the fertility data set whose size
is sufficient to assure the convergence of the algorithm, it does not
give better results than a single stepwise Cox model.

As far as the BNLS procedure is concerned, it seems not perform better
than a single survival tree on the two data sets, contrary to the
results found by Dannegger [6]. It can be explained by the fact that
it is difficult to tune a sensitive value for the complexity parameter
by cross-validation.  It may also result from the sufficient size of
the fertility data set.

\subsection{Selected variables in the final model}

The breast cancer data set was originally analyzed by Vijver
\textit{et al.} [31] who included in a multivariate Cox regression
model the clinical factors and a prognosis variable based on the gene
profile.  Indeed, the 70 genes expressions were not included as
independent covariates in the Cox model, as we do in our analysis.  We
notice that the procedures based on the Cox model selected no clinical
factors in the breast cancer data set. It can result from the fact
that the genes which were introduced in the model are those whose
expression is the most correlated positively or negatively with the
breast cancer survival duration among the multitude of genes
studied. Thus, these genes have a more important weight in the
multivariate analysis than clinical factors. In the BNLS and RSF
procedures, which take naturally into account interactions between
covariates, two clinical factors appears in the selected set of
covariates.  As it is not possible to include interactions in the Cox
model in a high-dimensional context, an alternative would be to force
the inclusion of some clinical factors clearly identified by the
 clinicians in the procedures based on the Cox model.

Regarding the fertility data set, we obtain different final models with
the procedures based on the Cox model and those based on survival
trees. However, all the selected covariates are relevant clinical
factors in reproductive issues. Indeed, the BSS, BNLS and RSF
procedures showed that female age is an important factor, with a
cut-off value of 34.5 years, which is similar to results observed in
other studies [34 -- 36].
Moreover, the infertility duration is selected by all the procedures,  with a cut-off  of
33.5 months in the BNLS method. It is not surprising as it is well-known that duration of
infertility is predictive in the occurrence of pregnancy [37]. It is also interesting to notice that the BLS, BSS
and BNLS procedures found a relationship between the live-birth and
male clinical factors whereas BRLS included very few covariates which
are essentially limited to female aspect and ART treatment. In most of
the publications about reproductive issue in infertile couples, authors
analysed the survival event of live-birth according only to female aspect
and type of ART treatment [38 -- 41]. 
However, we observe in our study that varicocelis
is included in  most of the procedures, which can be explained by the
fact that varicocelis has an impact on an impaired spermatogenesis
[42, 43]. The same
observation could be done for the testicular volume which
seems to be an important variable in the BNLS procedure, but with  less impact in the other final models.

\section{Concluding remarks}
 
As far as the breast cancer data set is concerned, the objective is
reasonably well achieved. The method based on randomized Lasso
selection combined with bootstrapping provides prediction error rates
very similar to those obtained by the RSF method, in association with
an easily interpretable final model for the medical field. Regarding
the procedure that combines bootstrap and survival trees, this method
especially allows to identify interactions between the relevant
covariates. 

On the contrary, the data set on fertility seems to be much more
difficult to analyze. The bootstrapping adapted to the various procedures based 
on the Cox model or on survival trees seems not improve the standard procedures (single Cox
stepwise selection or single survival tree). Moreover, the wide variations observed in the prediction error rates may be due to the weak predictive performance of the selected covariates. 

Finally, these results suggest that the Cox and tree based
procedures should be performed in a complementary way to identify the most relevant
covariates and provide to clinicians a stable and reliable model.
Each procedure  shows indeed a particular interest, either in  terms of its prediction performance, either in the selection of the relevant covariates.

\newpage
\section*{References}

\begin{enumerate}

	\item Cox DR. Regression models and life tables (with discussion). \textit{Journal of the Royal
Statistical Society: Series B} 1972; \textbf{34}:187--220.

	\item Harrell FE Jr, Lee KL, Califf RM, Pryor DB, Rosati RA. Regression
modelling strategies for improved prognostic prediction. \textit{Statistics in Medicine} 1984; \textbf{3}:143--152.

	\item Meinshausen N, Bühlmann P. Stability selection. \textit{Journal of the Royal Statistical Society: Series B} 2010,
 \textbf{72}:417--473.
 
	\item Bach F. Model-consistent sparse estimation through the bootstrap. \textit{Technical Report} 2009.

	\item Breiman L. Bagging predictors. \textit{Machine Learning} 1996; \textbf{24}:123--140.

	\item Dannegger F. Tree stability diagnostics and some remedies for instability. \textit{Statistics in Medicine} 2000;
\textbf{19}:475--491.

	\item Ruey-Hsia L. Instability of decision tree classification algorithms. \textit{PhD thesis} 2001.

	\item  Gey S, Poggi JM. Boosting and instability for regression trees. \textit{Computational
Statistics and Data Analysis}2006; \textbf{50}:533--550.

	\item Breiman L. Heuristics of instability and stabilization in model selection. \textit{Annals of
Statistics} 1996; \textbf{24}:2350--2383.

	\item Ishwaran H, Kogalur UB, Blackstone EH, Lauer MS. Random survival
forests. \textit{The Annals of Applied Statistics} 2008, \textbf{2}:841--860.

	\item van de Vijver MJ, He YD, van't Veer LJ, Dai H, Hart AA, Voskuil DW,
Schreiber GJ, Peterse JL, Roberts C, Marton MJ, Parrish M, Atsma D,Witteveen A, Glas A, Delahaye L, van der Velde T, Bartelink H, Rodenhuis S,
Rutgers ET, Friend SH, Bernards R. A gene-expression signature as a
predictor of survival in breast cancer. \textit{New England Journal of Medicine} 2002; \textbf{347}:1999--2009.

	\item Chen CH, George SL. The bootstrap and identification of prognostic factors
via cox's proportional hazards regression model. \textit{Statistics in Medicine} 1985; \textbf{4}:39--46.

	\item Sauerbrei W, Schumacher M. A bootstrap resampling procedure for model
building: application to the cox regression model. \textit{Statistics in Medicine} 1992; \textbf{11}:2093--2109.

	\item Tibshirani R. The lasso method for variable selection in the cox model. \textit{Statistics in Medicine} 1997;
\textbf{16}:385--395.

	\item Meinshausen N, Bühlmann P. High dimensional graphs and variable selection
with the lasso. \textit{Annals of Statistics} 2006; \textbf{34}:1436--1462.

	\item Freedman D. A remark on the difference between sampling with and without replacement.
\textit{Journal of the American Statistical Association} 1977; \textbf{72}:681.

	\item Breiman L. Classification and Regression Trees. \textit{Chapman and Hall/CRC} 1984.

	\item Gordon L, Olshen RA. Tree-structured survival analysis. \textit{Cancer Treatment Report} 1985;
\textbf{69}:1065--1069.

	\item Ciampi A, Thiffault J, Nakache JP, Asselain B. Stratification by stepwise
regression, correspondence analysis and recursive partition: A comparison of three
methods of analysis for survival data with covariates. \textit{Computational Statistics and
Data Analysis} 1986; \textbf{4}:185--204.

	\item Segal MR. Regression trees for censored data. \textit{Biometrics} 1988; \textbf{44}:35--48.

	\item LeBlanc M, Crowley J. Survival trees by goodness-of-split. \textit{Journal of the
American Statistical Association} 1993; \textbf{88}:457--467.

	\item Davis R, Anderson J. Exponential survival trees. \textit{Statistics in Medicine} 1989; \textbf{8}:947--
962.

	\item LeBlanc M, Crowley J. Relative risk trees for censored survival data. \textit{Biometrics} 1992;
\textbf{48}:411--425.

	\item Efron B. Censored data and the bootstrap. \textit{Journal of the American Statistical Association} 1981; \textbf{76}:312--319.

	\item Akritas MG. Bootstrapping the kaplan-meier estimator. \textit{Journal of the American Statistical Association} 1986; \textbf{81}:1032--1038.

	\item Freund Y, Schapire RE. A short introduction to boosting. \textit{Journal of Japanese
Society for Artificial Intelligence} 1999, \textbf{14}:771--780.

	\item Dietterich T. Ensemble methods in machine learning. \textit{In J. Kittler and F. Roli (Ed.)} First International Workshop on Multiple Classifier Systems, Lecture Notes in Computer Science. New York: Springer Verlag 2000; pp. 1--15.

	\item Radespiel-Troger M, Gefeller O, Rabenstein T, Hothorn T. Association between
split selection instability and predictive error in survival trees. \textit{Methods of Information in Medicine} 2006; \textbf{45}:548--556.

	\item Ishwaran H, Kogalur UB. Random survival forests for R. \textit{Rnews} 2007; \textbf{7}:25--31.

	\item Harrell FE, Davis CE. A new distribution-free quantile estimator. \textit{Biometrika} 1982; \textbf{69}:635--640.

	\item van de Vijver YD, He YD,van't Veer LJ, Dai H, Hart AA, Voskuil DW, 
Schreiber GJ, Peterse JL, Roberts C, Marton MJ, Parrish M, Atsma D, Witteveen A,
Glas A, Delahaye L, van der Velde T, Bartelink H, Rodenhuis S, Rutgers ET, 
Friend SH, Bernards R. A gene-expression signature as a predictor of survival in breast
cancer. \textit{New England of Journal of Medicine} 2002; 347:1999--2009.

	\item van 't Veer LJ, Dai H, van de Vijver MJ, He YD, Hart AA, Mao M, Peterse HL, van der Kooy K, Marton MJ,
Witteveen AT, Schreiber GJ, Kerkhoven RM, Roberts C, Linsley PS, Bernards R, Friend SH. Gene expression
profiling predicts clinical outcome of breast cancer. \textit{Nature} 2002; \textbf{415}:530--536.

	\item Omurlu IK, Ture M, Tokatli F. The comparisons of random survival forests
and cox regression analysis with simulation and an application related to breast
cancer. \textit{Expert Systems with Applications} 2009; \textbf{36}:8582--8588.

	\item de la Rochebrochard A, Thonneau P. Paternal age and maternal age are
risk factors for miscarriage; results of a multicentre european study. \textit{Human Reproduction} 2002;
\textbf{17}:1649--1656.

	\item Marquard K, Westphal LM, Milki AA, Lathi RB. Etiology of recurrent
pregnancy loss in women over the age of 35 years. \textit{Fertility and Sterility} 2009; \textbf{94}:1473--1477.

	\item Malizia BA, Hacker MR, Penzias AS. Cumulative live-birth rates after in
vitro fertilization. \textit{New England Journal of Medicine} 2009; \textbf{360}:236--243.

	\item Ducot B, Spira A, Feneux D, Jouannet P. Male factors and the likelihood
of pregnancy in infertile couples. II. study of clinical characteristics -- practical consequences.
\textit{International Journal of Andrology} 1988; \textbf{11}:395--404.

	\item Lintsen AM, Eijkemans MJ, Hunault CC, Bouwmans CA, Hakkaart L, Habbema JD, Braat DD.
 Predicting ongoing pregnancy chances after IVF and ICSI:
a national prospective study. \textit{Human Reproduction} 2007; \textbf{22}:2455--2462.

	\item Stolwijk AM, Wetzels AM, Braat DD. Cumulative probability of achieving
an ongoing pregnancy after in-vitro fertilization and intracytoplasmic sperm injection
according to a woman's age, subfertility diagnosis and primary or secondary
subfertility. \textit{Human Reproduction} 2000; \textbf{15}:203--209.

	\item Olivius K, Friden B, Lundin K, Bergh C. Cumulative probability of live birth
after three in vitro fertilization/intracytoplasmic sperm injection cycles. \textit{Fertility and Sterility} 2002;
\textbf{77}:505--510.

	\item Sharma V, Allgar V, Rajkhowa M. Factors influencing the cumulative conception
rate and discontinuation of in vitro fertilization treatment for infertility. \textit{Fertility and Sterility} 2002; \textbf{78}:40--46.

	\item Redmon JB, Carey P, Pryor JL. Varicocele: the most common cause of male factor infertility? \textit{Human Reproduction Update} 2002; \textbf{8}:53--58.

	\item World Health Organisation. The influence of varicocele on parameters of fertilityin a large group of men presenting to infertility clinics. \textit{Fertility and Sterility} 1992; \textbf{57}:1289--1293.

\end{enumerate}

\end{document}